\documentclass[reprint,eqsecnum,floats,aps,amsmath,amssymb,nofootinbib,prd,onecolumn, showpacs]{revtex4-2}

\usepackage{graphicx}
\usepackage{amsmath}
\usepackage{hyperref}
\usepackage{subfigure}
\usepackage{arydshln}
\usepackage{inputenc}
\usepackage{amssymb}

\begin{document}

\title{Hamiltonian formulation and loop quantization of a recent extension of the Kruskal spacetime}

\author{Beatriz Elizaga Navascu\'es}
\email{w.iac20060@kurenai.waseda.jp}
\affiliation{JSPS International Research Fellow, Department of Physics, Waseda University, 3-4-1 Okubo, Shinjuku-ku, 169-8555 Tokyo, Japan}
\author{Alejandro Garc\'ia-Quismondo}
\email{alejandro.garcia@iem.cfmac.csic.es}
\affiliation{Instituto de Estructura de la Materia, IEM-CSIC, Serrano 121, 28006 Madrid, Spain}
\author{Guillermo A. Mena Marug\'an}
\email{mena@iem.cfmac.csic.es}
\affiliation{Instituto de Estructura de la Materia, IEM-CSIC, Serrano 121, 28006 Madrid, Spain}

\begin{abstract}
We study the Hamiltonian formulation of the Ashtekar-Olmedo-Singh model for the description of the interior geometry of non-rotating, uncharged black holes. This model incorporates loop quantum effects through the introduction of two regularization parameters. We consider an extended phase space formalism proposed by the creators of the model that includes such parameters as configuration variables, constrained to be functions of the black hole mass. We generalize this restriction, allowing for an off-shell phase space dependence. We then introduce a gauge fixing procedure and reduce the system, proving that the reduced symplectic structure cannot reproduce the standard relativistic one in terms of the densitized triad and the Ashtekar-Barbero connection. Actually, the reduced structure precisely compensates the modifications that arise in the Hamilton equations when the regularization parameters are treated as phase space functions, rather than as numbers, attaining a consistent Hamiltonian derivation of the dynamics. We then choose the extended phase space formalism as starting point to address the loop quantization of the model. Taking the definition of certain geometric operators as the only basic ingredient and adopting prescriptions that have proven successful in loop quantum cosmology, we construct a polymer representation of all the constraints and deduce the formal expression of the physical states, assuming reasonable spectral properties for the constraint operators. The physical states turn out to be characterized by a wave function of the black hole mass with support on a very specific set. We finally discuss conditions that guarantee the existence of physical states in the region of large black hole masses. This is a first step in the development of a new loop quantum theory of black holes. 

\end{abstract}

\maketitle

\section{Introduction}

Among the different approaches for the quantization of general relativity, loop quantum gravity (LQG) \cite{ashlewlqg,lqgThiemann} stands out as a promising candidate, leading to predictions that might eventually make contact with observations. For example, the application of LQG techniques to the description of the very early Universe, in the discipline known as loop quantum cosmology (LQC) \cite{abl,lqc2,lqc3,hyb-pert1,AAN1,AAN2,Bojo2,hyb-pert4,Hybridreview}, is an active field that might unveil traces of quantum gravitational phenomena in the cosmic microwave background \cite{Bojo1,Ivan,Edward,hybpred1,hybpred2,AshNe,hybothers,JCAP}. Another appealing scenario for testing the quantum nature of gravity is black hole physics, especially as we are witnessing the dawn of gravitational wave astronomy.

Over the last decades, there have been many proposals to apply the quantization program of LQG to spacetimes that correspond to black holes in general relativity \cite{ABbh,Modestobh,CKbh,DJSbh,BVbh,Chioubh,CGPbh1,SKbh,BKdBbh,GOPbh,DJSbh,HRbh,CGOPbh,CSbh,OSSbh,Jerobh,YKSbh,BCDHRbh,ABP,Bojobh,GOPbheff,Edbh1,Edbh2}. The case of spherically symmetric spacetimes is especially interesting, owing to their simplicity but yet rich physical properties. In this context, a few years ago Ashtekar, Olmedo, and Singh (AOS) proposed a loop quantum extension of the Kruskal spacetime which has received a fair amount of attention \cite{AOS1,AOS2,AOg}. This model incorporates quantum effects arising from LQC in the description of the interior of the Schwarzschild black hole, which behaves as an anisotropic cosmology. The resulting geometry is then smoothly extended to the exterior. As a consequence of the modifications based on LQC, the classical singularity in the interior is replaced with a transition surface that connects a trapped region with an anti-trapped one. Remarkably, the curvature invariants remain finite throughout the whole spacetime. Furthermore, in contrast with previous approaches to the loop quantum description of spherically symmetric models \cite{BVbh,Chioubh,CSbh,OSSbh,Jerobh}, the AOS model displays local quantum gravity effects near the horizons and in the exterior regions that are controllably small (for macroscopic black holes).

A fundamental difference between the construction of the AOS model and other previous LQC descriptions of black holes resides in the procedure followed to regularize the classical Hamiltonian \cite{AOS2}. In the formulation of general relativity in terms of gauge connections and densitized triads that constitutes the basis of LQG, the Hamiltonian depends on the curvature of the Ashtekar-Barbero connection \cite{ashlewlqg,lqgThiemann}. This curvature is not well defined as an operator in LQG, owing to the non-continuity of the quantum representation provided for the geometry \cite{gang,Thiemannreg}. This issue is inherited by LQC, so that one needs to prescribe a regularized version of the curvature prior to its quantization. This is usually done in terms of some parameters of quantum origin, which are related to the minimum non-zero eigenvalue allowed for the area operator in LQG \cite{abl,lqc2,lqc3}. The classical Hamiltonian is recovered in the limit in which this area gap tends to zero. The distinctive feature of the AOS formulation of the black hole interior is that these {\sl quantum parameters} are chosen to be specific functions of the mass of the black hole itself. Such a choice is the primary reason behind the appealing properties of the resulting spacetime geometry \cite{AOS1,AOS2,ABG}.

The regularization of the Hamiltonian chosen in the original works of the AOS model, leading to the extended Kruskal spacetime briefly described above, has been an active subject of debate \cite{norbert1,norbert2,norbert3,norbert4,GQMbh,AG2} (see Refs. \cite{Mariam,Bojocov} for criticisms on other features of the model). The reason is that, whereas the quantum parameters that are introduced for the regularization are chosen as functions of the black hole mass, their treatment in the calculation of the dynamical equations casts shadows on the correct derivation of the AOS solution. Indeed, that mass is in fact a constant of motion of the system and, therefore, it is a function on phase space. Thus, if the equations of motion follow from the regularized Hamiltonian, the phase space dependence of the mass should be taken into account when deducing them. On the contrary, the dynamical equations that describe the AOS geometry can only be obtained from this Hamiltonian if the quantum parameters are handled as pure constants in the calculations, making them equal to functions of the mass only at the very end, namely on-shell. 

In order to motivate the dynamical equations for the black hole interior from a canonical approach, the authors of the AOS model have argued that they can be derived if one suitably extends the phase space to include the quantum parameters as canonical variables \cite{AOS2}. In this extended formulation, these parameters are subject to constraints that dictate their relation with the mass as given functions on phase space. Nonetheless, the symplectic relation between the reduction of this extended phase space (after imposing the constraints) and the original phase space of the black hole interior is unclear. Understanding this relation is of major importance if one wishes to go beyond the study of a spacetime geometry with loop corrections, and explore a canonical quantum description of the black hole. In fact, the AOS model is conceived to portrait the semiclassical behavior of certain states in the loop quantization of non-rotating black holes. This is precisely what happens with the cosmological dynamics in LQC from which it draws inspiration, where certain families of Gaussian states describe bouncing universes within an effective dynamics \cite{APS}. However, that this is actually the case for black hole spacetimes is just an assumption at present. A viable quantum theory for the black hole interior is needed to support that the AOS geometry is a manifestation of quantum gravity phenomena. The construction of this theory should be founded on a Hamiltonian formulation of the system which, drawing a parallelism with LQC, ideally would lead to the AOS model after a suitable regularization.

With this perspective, the aim of this work is twofold. On the one hand, we want to clarify how the extended phase space proposed by the authors of the AOS model relates, after reduction, to the phase space of a Kantowski-Sachs cosmology in general relativity \cite{KS1,KS2,KS3}. This anisotropic spacetime describes the interior of the Schwarzschild black hole and, thus, it has traditionally served as the starting point for the study of the interior region of non-rotating black holes in LQC \cite{ABbh,Modestobh,BVbh,Chioubh,CGPbh1,BKdBbh,CSbh,DJSbh,OSSbh,Jerobh}. As we will see, the Poisson algebra of the connection and triad variables that describe the geometry differs greatly in the two formalisms under consideration. The variables are canonical in the case of Kantowski-Sachs, but not really in the extended phase space formalism (after reduction). This difference is precisely what makes possible that, using the symplectic structure inherited from the extended phase space, one can consistently derive the dynamical equations of the AOS model starting from the (regularized) Hamiltonian in Kantowski-Sachs, with the quantum parameters fixed as off-shell functions of the black hole mass. One is then inclined to believe that, if the AOS geometry is to be recovered effectively from a genuine quantum model of black hole spacetimes, then this quantum model should be based on the Hamiltonian formulation of the extended Kantowski-Sachs phase space. These considerations lead to the second purpose of this work, namely, paving the road to the loop quantization of the black hole (interior). We will argue in favor of using the extended Kantowski-Sachs model for its quantum description, and then proceed to outline the main steps for its quantization. The Hamiltonian of the system is a linear combination of constraints: the relativistic one, identical to the one found in Kantowski-Sachs cosmologies, and the ones that implement the relation between the quantum parameters and the black hole mass. We will show how to construct a quantum representation for these constraints following well known techniques of LQC and LQG. Finally, we will formally show how physical states (namely, those annihilated by the constraints) look like, and discuss some of their properties. 

The content of this paper is organized as follows. In Sec. \ref{sec1}, we summarize the AOS model of a black hole interior. In order to make this paper self-contained, we briefly introduce the canonical description of Kantowski-Sachs cosmologies in terms of Ashtekar-Barbero variables. Then, we formulate the effective Hamiltonian and equations of motion of the AOS model, noticing some caveats in their relation. Finally, we summarize the extended phase space formulation used in Ref. \cite{AOS2} to motivate these equations of motion from a Hamiltonian perspective. In Sec. \ref{sec2}, we consider a natural generalization of the extended phase space, that takes into account the on-shell indistinguishability of two different identifications of the black hole mass as a phase space function. The aim of this section is to clarify the relation between phase space reductions leading to the AOS model and the Kantowski-Sachs cosmologies. For this, we first study how such reductions affect the symplectic algebra of connection and triad variables. We explicitly derive this algebra and show that it is inequivalent to the canonical one found in the case of the Kantowski-Sachs cosmologies. This inequivalence actually turns out to allow for a consistent derivation of the AOS model in the reduced phase space. In Sec. \ref{sec3}, we address the loop quantization of the extended phase space formalism. We borrow techniques from LQC to find a representation of the holonomy-flux algebra, and use it to construct a quantum representation of the constraints of the system after their regularization (according to the usual strategies in LQG). Finally, in Sec. \ref{sec:phys} we formally characterize the physical states annihilated by these constraints and discuss conditions to consistently recover a reasonable sector of large black hole masses. In Sec. \ref{conclu}, we summarize our results and comment on the outlook of our research. Two appendices with details are included. Throughout this article, we adopt geometrical natural units, setting the speed of light, Planck's reduced constant, and Newton's gravitational constant equal to one.

\section{The AOS model}\label{sec1}

Let us start by considering homogeneous but anisotropic classical spacetimes that exhibit spherical symmetry. It is well known that their geometry can be described by a Kantowski-Sachs metric \cite{KS1,KS2,KS3}. Furthermore, in suitable coordinates, they can model the interior of the Schwarzschild black hole \cite{ABbh}. The topology of the spatial hypersurfaces is given by $I\times S^2$, where $I=(0,L_o)$ and $L_o$ is a fiducial coordinate length\footnote{The consideration of the interval $I$ instead of the real line avoids possible infrared divergences in the Hamiltonian formulation. Physical quantities must have a well defined limit when $L_o$ tends to infinity.}. The components of the densitized triad and Ashtekar-Barbero connection can be respectively written as
\begin{align}\label{KStriad}
E_i^{\alpha}  \partial_{\alpha} =\delta_i^3 p_c \sin\theta \partial_x +\delta_i^2 \frac{p_b}{L_o} \sin\theta \partial_\theta -\delta_i^1 \frac{p_b}{L_o}\partial_\phi,\\ \label{AB}
A^i_{\alpha} \text{d}x^{\alpha}=\delta^i_3\frac{c}{L_o} \text{d}x +\delta^i_2 b \text{d}\theta -\delta^i_1 b \sin\theta \text{d}\phi +\delta^i_3 \cos\theta \text{d}\phi,
\end{align}
where $(x,\theta,\phi)$ is a set of coordinates adapted to the spatial isometries, with $x\in I$, $\theta\in [0,\pi)$, and $\phi\in [0,2\pi)$. Throughout this paper, we adopt a convention such that letters from the beginning of the Greek alphabet denote spatial indices on tensor fields and take values in the set $(x,\theta,\phi)$, whereas $i,j,...=1,2,3$ represent internal $su(2)$ indices. The variables $p_b$, $p_c$, $b$, and $c$ are functions of the coordinate time of the system $\tau$ and they codify the dynamical information about the triad and connection components. In terms of these and the lapse function $N$, the spacetime line element reads
\begin{align}\label{KSmetric}
ds^2=-N(\tau)^2 \text{d}\tau^2 +\frac{p_b^2(\tau)}{L_o^2 |p_c(\tau)|}\text{d}x^2+|p_c(\tau)|\left(\text{d}\theta^2+\sin^2\theta \text{d}\phi^2\right).
\end{align}

In general relativity, our variables satisfy a canonical algebra with the following non-zero Poisson brackets:
\begin{align}\label{KSPoisson}
\{b,p_b\}=\gamma, \qquad \{c,p_c\}=2\gamma,
\end{align}
where $\gamma\simeq 0.2375$ is the commonly used value of the Immirzi parameter in LQG \cite{ashlewlqg,lqgThiemann}. In addition, the Hamiltonian $H_{\text{KS}}[N]$ for the considered Kantowski-Sachs cosmologies can be expressed as
\begin{align}\label{KSHam}
H_{\text{KS}}[N]=N L_o\frac{b}{\gamma\sqrt{|p_c|}}\left(O^{\text{KS}}_b-O^{\text{KS}}_c\right),\qquad O^{\text{KS}}_b=-\frac{p_b}{2\gamma L_o}\left(b+\frac{\gamma^2}{b}\right),\qquad O^{\text{KS}}_c=\frac{cp_c}{\gamma L_o}.
\end{align}
In the classical theory, with the choice of lapse $N=\gamma\sqrt{|p_c|}/b$, the phase space sectors coordinatized by $(b,p_b)$ and $(c,p_c)$ are dynamically decoupled. Furthermore, the partial Hamiltonians $O^{\text{KS}}_b$ and $O^{\text{KS}}_c$, which generate the respective dynamics of these sectors with a suitable choice of time, are constants of motion and equal to each other on-shell. When the Kantowski-Sachs geometry is used to describe the interior region of the Schwarzschild black hole in general relativity, the absolute value of the resulting constant of motion turns out to be the ADM mass of the spacetime.

\subsection{The AOS dynamics}

In homogeneous and isotropic LQC, well established results show that a wide class of physical states are peaked at trajectories for the phase space variables that follow a dynamics generated by an effective Hamiltonian \cite{lqc2,lqc3,APS}. Remarkably, it turns out that this effective Hamiltonian can be obtained by replacing the connection variable $\zeta$ in that model by $\sin(\bar\mu \zeta)/\bar\mu$ \cite{Taveras}, where $\bar\mu$ is a (phase space dependent) regularization parameter of quantum origin. Specifically, $\bar\mu=\sqrt{\Delta/|p_\zeta|}$, where $p_\zeta$ is the canonically conjugate momentum of $\zeta$ and $\Delta$ is the minimum non-zero area allowed by the spectrum of the area operator in LQG \cite{APS}. Motivated by these results, the AOS model assumes an effective dynamical description of the black hole interior that, with the choice of lapse described in the preceding paragraph, is generated by the effective Hamiltonian \cite{AOS1,AOS2}
\begin{align}\label{effectiveH}
H^{\text{eff}}_{\text{AOS}}=L_o\left({O}_b-{O}_c\right), \qquad {O}_b=-\frac{p_b}{2\gamma L_o}\left[\frac{\sin(\delta_b b)}{\delta_b}+\frac{\gamma^2 \delta_b}{\sin(\delta_b b)}\right],\qquad {O}_c=\frac{\sin(\delta_c c)}{\gamma L_o \delta_c}p_c,
\end{align}
where $\delta_b$ and $\delta_c$ are the (real) quantum parameters of the model. Clearly, if these parameters are genuine constant numbers or functions of ${O}_b$ and/or ${O}_c$, we find ourselves again in a situation where these two partial Hamiltonians are equal on-shell to one and the same constant of motion, which we will call $m$.

In the original formulation of the AOS model, it is claimed that the the following equations of motion for the connection and triad variables follow from the above Hamiltonian \cite{AOS1,AOS2}:
\begin{align}\label{AOS1}
\dot{b}=-\frac{1}{2}\left[\frac{\sin(\delta_b b)}{\delta_b}+\frac{\gamma^2 \delta_b}{\sin(\delta_b b)}\right],\qquad \dot{c}=-2\frac{\sin(\delta_c c)}{\delta_c},\\ \label{AOS2}
\dot{p}_b=\frac{p_b}{2}\cos(\delta_b b)\left(1-\frac{\gamma^2 \delta_b^2}{\sin^2(\delta_b b)}\right),\qquad \dot{p}_c=2p_c \cos(\delta_cc),
\end{align}
for a large class of quantum parameters whose behavior is fixed so that, in the limit of large $|m|$,
\begin{align}\label{deltas}
\delta_b=\left(\frac{\sqrt{\Delta}}{\sqrt{2\pi}\gamma^2 m}\right)^{1/3},\qquad L_o\delta_c=\frac{1}{2}\left(\frac{\gamma\Delta^2}{4\pi^2 m}\right)^{1/3}.
\end{align}

The AOS black hole interior geometry is attained by solving Eqs. \eqref{AOS1} and \eqref{AOS2}. However, the actual Hamiltonian derivation of these dynamical equations is rather obscure. Indeed, using the symplectic structure of the Kantowski-Sachs cosmologies [see Eq. \eqref{KSPoisson}], the only way to derive these equations from the effective Hamiltonian $H^{\text{eff}}_{\text{AOS}}$ is to treat the quantum parameters $\delta_b$ and $\delta_c$ as constant numbers. Nevertheless, they are finally fixed as functions of the mass $m$, which, we recall, is in fact a constant of motion: it takes a different value on each of the dynamical solutions. Because of this non-trivial phase space dependence, functions of $m$ do not behave as constant numbers under Poisson brackets. The realization of this tension has led several authors to question the fundamental relation between the effective Hamiltonian and the dynamics of the AOS model \cite{norbert1,GQMbh,AG2}. 

\subsection{Extended phase space}\label{subsec:IIB}

In order to attain a Hamiltonian derivation of the dynamical equations of the AOS model consistent with the identification of the quantum parameters as constants of motion, an extension of the Kantowski-Sachs phase space $\Gamma_{\text{KS}}$ and of its dynamics has been proposed in Ref. \cite{AOS2}. In that work, an extended phase space $\Gamma_{\text{ext}}$ of dimension eight is defined in the first place. In addition to the pairs $(b,p_b)$ and $(c,p_c)$, which satisfy the canonical algebra \eqref{KSPoisson}, two new pairs $(\delta_b,p_{\delta_b})$ and $(\delta_c,p_{\delta_c})$ are introduced, their only non-zero Poisson brackets being  
\begin{align}
\{\delta_b,p_{\delta_b}\}=1, \qquad \{\delta_c,p_{\delta_c}\}=1.
\end{align}
The Hamiltonian on this extended phase space has the form 
\begin{align}\label{hamyum}
\underline{N}H^{\text{eff}}_{\text{AOS}}+\lambda_b \Phi_b +\lambda_c \Phi_c,
\end{align}
where $\underline{N}$, $\lambda_b$, and $\lambda_c$ are non-dynamical Lagrange multipliers, and the constraints $\Phi_b$ and $\Phi_c$ take the expressions 
\begin{align}
\Phi_b={O}_b - F_b (\delta_b), \qquad \Phi_c={O}_c - F_c (\delta_c),
\end{align}
for a certain pair of (at least) $\mathcal{C}^1$ functions, $F_b$ and $F_c$. In principle, the only restriction on them [arising from Eq. \eqref{deltas}] is that they must behave as
\begin{align}
F_b(\delta_b)=\frac{\sqrt{\Delta}}{\sqrt{2\pi}\gamma^2 \delta_b^3},\qquad F_c (\delta_c)=\frac{\gamma\Delta^2}{32\pi^2 (L_o \delta_c)^3},
\end{align}
at dominant order in the limit of small absolute values of $\delta_b$ and $\delta_c$. Clearly, all of the constraints in this Hamiltonian commute with each other under Poisson brackets, so 
they form a first-class set and can be interpreted as the generators of symmetries of the system. Under the choice of Lagrange multipliers $\underline{N}=1$ and $\lambda_c=\lambda_b=0$, the evolution generated by this Hamiltonian on the subspace of $\Gamma_{\text{ext}}$ coordinatized by $(b,p_b)$ and $(c,p_c)$ is ruled by the AOS dynamics \eqref{AOS1}-\eqref{deltas}. These equations can be equivalently obtained if one i) suitably fixes the freedom associated with the constraints $\Phi_b$ and $\Phi_c$ to eliminate the degrees of freedom associated with the pairs $(\delta_b,p_{\delta_b})$ and $(\delta_c,p_{\delta_c})$, and ii) computes the dynamics on the phase space $\bar{\Gamma}_{\text{ext}}$ resulting from the reduction of the system, provided that the choice of gauge leads to $\lambda_b=\lambda_c=0$. A particular class of such gauge choices for $(\delta_b,p_{\delta_b})$ and $(\delta_c,p_{\delta_c})$ is considered in Ref. \cite{AOS2}, where the AOS equations for the black hole interior are explicitly obtained after the gauge fixing. In view of this procedure, it seems natural to ask what kind of relation actually exists between $\bar{\Gamma}_{\text{ext}}$ and $\Gamma_{\text{KS}}$. This question is of key importance to understand how the AOS dynamical equations can arise canonically from an effective LQC description of Kantowski-Sachs cosmologies.

\section{Reduction of the extended phase space}\label{sec2}

In this section we will consider reductions of the extended phase space $\Gamma_{\text{ext}}$ that remove the degrees of freedom $(\delta_b,p_{\delta_b})$ and $(\delta_c,p_{\delta_c})$ and lead to vanishing Lagrange multipliers $\lambda_b$ and $\lambda_c$, so that the AOS equations may rule the reduced dynamics. Then, we will study the relation between the result of this reduction and the Kantowski-Sachs phase space $\Gamma_{\text{KS}}$. 

With the aim of generalizing previous studies while respecting the good physical properties of the system, we will carry out our analysis with a Hamiltonian different from \eqref{hamyum}, but equally valid on $\Gamma_{\text{ext}}$. Indeed, let us notice that the constant of motion $m$ of the AOS model is indistinguishable from both ${O}_b$ and ${O}_c$ on-shell \cite{GQMbh,AG2}. Therefore, we can more generally consider the Hamiltonian
\begin{align}\label{exHeff}
H^{\text{eff}}_{\text{ext}}=\underline{N}H^{\text{eff}}_{\text{AOS}}+\lambda_b \Psi_b +\lambda_c \Psi_c,
\end{align}
where the constraints $\Psi_b$ and $\Psi_c$ associated with the quantum parameters take the form
\begin{align}\label{AOSconstraints}
\Psi_b=K_b\left({O}_b,{O}_c\right) - \delta_b, \qquad \Psi_c=K_c\left({O}_b,{O}_c\right) - \delta_c,
\end{align}
for a certain pair of sufficiently smooth functions, $K_b$ and $K_c$. {\sl A priori}, the only restriction on these functions is that, at dominant order in the limit of large $|m|$, 
\begin{align}\label{deltasext}
K_b(m,m)=\left(\frac{\sqrt{\Delta}}{\sqrt{2\pi}\gamma^2 m}\right)^{1/3},\qquad K_c(m,m)=\frac{1}{2 L_o}\left(\frac{\gamma\Delta^2}{4\pi^2 m}\right)^{1/3},
\end{align}
as it is required to correctly generate the equations of motion of the AOS model under the choice of Lagrange multipliers given by $\underline{N}=1$ and $\lambda_c=\lambda_b=0$. Once again, it is clear that all of the constraints in $H^{\text{eff}}_{\text{ext}}$ form a first-class set which generates the symmetries of the system.

\subsection{Gauge fixing procedure: Dirac algebra} \label{subsec:gaugefixing}

In order to remove $\delta_b$ and $\delta_c$ as degrees of freedom, one can add to the constraints $\Psi_b$ and $\Psi_c$ conditions that fix the gauge associated with the canonical momenta $p_{\delta_b}$ and $p_{\delta_c}$. Following the strategy employed in Ref. \cite{AOS2}, we introduce the gauge fixing conditions $\chi_b=\chi_c=0$, with
\begin{align}
\chi_b=P_{\delta_b}-G_b({O}_b,{O}_c),\qquad \chi_c=P_{\delta_c}-G_c({O}_b,{O}_c),
\end{align}
where $G_b$ and $G_c$ are two sufficiently smooth functions. Here, $P_{\delta_b}$ and $P_{\delta_c}$ are suitably defined momenta that are canonically conjugate to $\delta_b$ and $\delta_c$, respectively, and constructed so that they Poisson commute with the partial Hamiltonians ${O}_b$ and ${O}_c$. Actually, one can change variables from $b$, $c$, and their momenta to ${O}_b$, ${O}_c$, and suitable momenta $P_b$ and $P_c$, obtaining a canonical set for the extended phase space together with $(\delta_b,P_{\delta_b},\delta_c,P_{\delta_c})$. Further details on this canonical transformation, including the expressions of the new momenta, can be found in Appendix \ref{appendix:Pdeltas}, although we encourage the reader to consult Ref. \cite{AOS2} for a complete description. Let us mention here only the properties of $P_{\delta_b}$ and $P_{\delta_c}$ that are relevant for the present discussion, namely that these momenta depend exclusively on the respective $b$ and $c$ sectors and differ from the original momenta $p_{\delta_b}$ and $p_{\delta_c}$ in terms independent of them:
\begin{align}\label{newpdelta}
P_{\delta_b}=P_{\delta_b}(b,p_b,\delta_b,p_{\delta_b}),\quad \frac{\partial P_{\delta_b}}{\partial p_{\delta_b}}=1,\qquad P_{\delta_c}=P_{\delta_c}(c,p_c,\delta_c,p_{\delta_c}),\quad \frac{\partial P_{\delta_c}}{\partial p_{\delta_c}}=1.
\end{align}
The constraints $\Psi_b$, $\Psi_c$, $\chi_b$, and $\chi_c$ form a second-class set, indicating that our conditions provide a good gauge fixing. Indeed, the Poisson algebra of these constraints reads
\begin{align}\label{transg}
\{\chi_b,\Psi_b\}=1,\qquad \{\chi_c,\Psi_c\}=1,\qquad \{\chi_b,\Psi_c\}=0,\qquad \{\chi_c,\Psi_b\}=0,\qquad \{\chi_b,\chi_c\}=0, \qquad \{\Psi_b,\Psi_c\}=0.
\end{align}
In addition, for the gauge fixing to be well posed, it must be stable under the dynamical evolution generated by the Hamiltonian $H^{\text{eff}}_{\text{ext}}$. This requires that the Lagrange multipliers $\lambda_b$ and $\lambda_c$ can be fixed so that the Poisson brackets of the four considered constraints with the total Hamiltonian vanish on our gauge fixing section. A straightforward calculation shows that this is indeed the case if and only if $\lambda_b=\lambda_c=0$.

After this gauge fixing, the pairs $(\delta_b,P_{\delta_b})$ and $(\delta_c,P_{\delta_c})$ become functions of the rest of phase space. In this sense, the eight-dimensional extended phase space $\Gamma_{\text{ext}}$ is reduced to a four dimensional one, $\bar{\Gamma}_{\text{ext}}$. The canonical algebra of functions on $\bar{\Gamma}_{\text{ext}}$ is obtained from the pull back of the symplectic structure on $\Gamma_{\text{ext}}$. An explicit way in which one can derive this canonical algebra on the reduced phase space is by considering the passage from Poisson to Dirac brackets. If $\{\cdot,\cdot\}$ are the Poisson brackets of the phase space subject to the set of second-class constraints $(\varphi_1,...,\varphi_{4})=(\Psi_b,\Psi_c,\chi_b,\chi_c)$, the Dirac bracket $\{\cdot,\cdot\}_{D}$ between two functions $f$ and $g$ on the reduced phase space where these constraints have been implemented is defined as \cite{Dirac}
\begin{align}\label{Dirac}
\{f,g\}_D=\{f,g\}-\sum_{\mu,\nu=1}^{4}\{f,\varphi_\mu\}(M^{-1})_{\mu\nu}\{\varphi_{\nu},g\},
\end{align}
where $M$ is the $4\times 4$ matrix with elements given by $(M)_{\mu\nu}=\{\varphi_\mu,\varphi_\nu\}$. In the case under consideration, we simply have
\begin{align}\label{M}
M^{-1}=\begin{pmatrix}
0_2 && \mathbb{I}_2 \\[0.5em]
-\mathbb{I}_2 && 0_2\\[0.5em]
\end{pmatrix}.
\end{align}
Here, $0_2$ and $\mathbb{I}_2$ are the two-dimensional null and identity matrices, respectively.

After reducing the phase space with our gauge fixing conditions, the Dirac algebra of the variables ${O}_b$, $P_b$, ${O}_c$, and $P_c$ on $\bar{\Gamma}_{\text{ext}}$ can be easily computed to be
\begin{align}
\{{O}_b,P_b\}_D=1,\quad \{{O}_c,P_c\}_D=1,\quad \{{O}_b,{O}_c\}_D=0,\quad \{P_b,P_c\}_D=\sum_{a=b,c}\left(\frac{\partial G_{a}}{\partial {O}_b}\frac{\partial K_{a}}{\partial {O}_c}-\frac{\partial G_{a}}{\partial {O}_c}\frac{\partial K_{a}}{\partial {O}_b}\right).
\end{align}
Then, it is clear that, for general choices of gauge (namely, general choices of functions $G_{a}$ and $K_{a}$), the resulting algebra ceases to be canonical on the reduced phase space, because the last bracket is not identically zero.

\subsection{Inequivalence with Kantowski-Sachs cosmologies}

In the original motivation for an extended phase space formulation of the AOS model, it is assumed that a choice of gauge such that $\{P_b,P_c\}_D=0$ leads to a reduced phase space $\bar{\Gamma}_{\text{ext}}$ that is symplectomorphic to the phase space $\Gamma_{\text{KS}}$ of the Kantowski-Sachs cosmologies \cite{AOS2}. It is in this sense that the AOS dynamical equations are understood as arising from a Hamiltonian description of the system. However, the implicit dependence of the quantum parameters $\delta_b$ and $\delta_c$ appearing in the partial Hamiltonians ${O}_b$ and ${O}_c$ on these same quantities obscures the validity of such an assumption. In fact, in the following, we will explicitly show that $\bar{\Gamma}_{\text{ext}}$ and $\Gamma_{\text{KS}}$ describe inequivalent phase spaces, even if one fixes the gauge such that $\{P_b,P_c\}_D=0$. We will do so by checking the Dirac algebra that the connection and triad variables $b$, $p_b$, $c$, and $p_c$ satisfy in $\bar{\Gamma}_{\text{ext}}$. We will prove that they fail to form canonical pairs, unlike what happens in $\Gamma_{\text{KS}}$.

Let us first restrict our analysis to any choice of gauge that guarantees the commutativity of $P_b$ and $P_c$ under Dirac brackets, namely to functions $G_b$ and $G_c$ that satisfy
\begin{align}
\frac{\partial G_b}{\partial {O}_b}\frac{\partial K_b}{\partial {O}_c}-\frac{\partial G_b}{\partial {O}_c}\frac{\partial K_b}{\partial {O}_b}=\frac{\partial G_c}{\partial {O}_c}\frac{\partial K_c}{\partial {O}_b}-\frac{\partial G_c}{\partial {O}_b}\frac{\partial K_c}{\partial {O}_c}.
\end{align}
A direct application of Eqs. \eqref{Dirac} and \eqref{M} then reveals that the variables $b$, $p_b$, $c$, and $p_c$ have the following non-zero Dirac brackets in $\bar{\Gamma}_{\text{ext}}$:
\begin{align}
\{b,p_b\}_D=\gamma\left(1-\frac{\partial K_b}{\partial {O}_b}\frac{\partial {O}_b}{\partial \delta_b}\right),\qquad \{c,p_c\}_D=2\gamma\left(1-\frac{\partial K_c}{\partial {O}_c}\frac{\partial {O}_c}{\partial\delta_c}\right),
\end{align}
\begin{align}\label{diracbc}
\{b,c\}_D=2\gamma^2\left(\frac{\partial P_{\delta_c}}{\partial p_c}\frac{\partial K_c}{\partial {O}_b}\frac{\partial {O}_b}{\partial p_b}-\frac{\partial P_{\delta_b}}{\partial p_b}\frac{\partial K_b}{\partial {O}_c}\frac{\partial {O}_c}{\partial p_c}\right), 
\end{align}
\begin{align}
\{b,p_c\}_D=2\gamma^2\left(\frac{\partial P_{\delta_b}}{\partial p_b}\frac{\partial K_b}{\partial {O}_c}\frac{\partial {O}_c}{\partial c}-\frac{\partial P_{\delta_c}}{\partial c}\frac{\partial K_c}{\partial {O}_b}\frac{\partial {O}_b}{\partial p_b}\right), 
\end{align}
\begin{align}
\{c,p_b\}_D=2\gamma^2\left(\frac{\partial P_{\delta_c}}{\partial p_c}\frac{\partial K_c}{\partial {O}_b}\frac{\partial {O}_b}{\partial b}-\frac{\partial P_{\delta_b}}{\partial b}\frac{\partial K_b}{\partial {O}_c}\frac{\partial {O}_c}{\partial p_c}\right),
\end{align}
\begin{align}\label{diracpcpb}
\{p_b,p_c\}_D=2\gamma^2\left(\frac{\partial P_{\delta_c}}{\partial c}\frac{\partial K_c}{\partial {O}_b}\frac{\partial {O}_b}{\partial b}-\frac{\partial P_{\delta_b}}{\partial b}\frac{\partial K_b}{\partial {O}_c}\frac{\partial {O}_c}{\partial c}\right),
\end{align}
as one can readily check using that $P_{\delta_b}$ and $P_{\delta_c}$ exhibit the properties displayed in Eq. \eqref{newpdelta} and that they Poisson commute by construction with ${O}_b$ and ${O}_c$ in the extended phase space $\Gamma_{\text{ext}}$. Clearly, the Dirac brackets $\{b,p_b\}_D$ and $\{c,p_c\}_D$ are not constant for general choices of functions $K_b$ and $K_c$. In particular, this happens if one defines $K_b$ and $K_c$ to depend only on ${O}_b$ and ${O}_c$, respectively, as it is done in the original Ref. \cite{AOS2} of the AOS model (see Sec. \ref{subsec:IIB}). In that case, it is true that the rest of Dirac brackets \eqref{diracbc}-\eqref{diracpcpb} vanish, but the connection variables still do not have canonical brackets with the triad ones. In general, the non-trivial cases in which $\{b,p_b\}_D$ and $\{c,p_c\}_D$ are constant occur when $K_b$ and $K_c$ are respectively independent of ${O}_b$ and ${O}_c$. Then, however, it is not hard to convince oneself that the rest of Dirac brackets cannot be all zero. Finally, the trivial situation with constant functions $K_b$ and $K_c$ is excluded by the very construction of the AOS model. We can thus conclude that the connection and triad variables can never form a canonical set in $\bar{\Gamma}_{\text{ext}}$. As an immediate consequence, the reduced phase space is generally inequivalent to the one that describes the Kantowski-Sachs cosmologies.

The change in the canonical structure of $\bar{\Gamma}_{\text{ext}}$ with respect to that of $\Gamma_{\text{KS}}$ can be used to reconcile the AOS dynamical equations with an effective Hamiltonian $\underline{N}H^{\text{eff}}_{\text{AOS}}$ where the quantum parameters $\delta_b$ and $\delta_c$ are phase space functions that behave as constants of motion on solutions. Indeed, after imposing the constraints $\Psi_b=\Psi_c=0$ associated with these parameters, the Dirac brackets between the variables $b$, $p_b$, and the partial Hamiltonians ${O}_b$ and ${O}_c$ can be written as
\begin{align}\label{dynb}
&\{b,O_b\}_D=\frac{1}{\beta}\bigg[\frac{\partial {O}_b}{\partial p_b}\left(1-\frac{\partial K_c}{\partial {O}_c}\frac{\partial {O}_c}{\partial \delta_c}\right)\{b,p_b\}_D+\frac{\partial K_b}{\partial {O}_c}\frac{\partial {O}_b}{\partial \delta_b}\left(\frac{\partial {O}_c}{\partial c}\{b,c\}_D+\frac{\partial {O}_c}{\partial p_c}\{b,p_c\}_D\right)\bigg],\nonumber \\& \{b,O_c\}_D=\frac{1}{\beta}\bigg[\frac{\partial {O}_b}{\partial p_b}\frac{\partial K_c}{\partial {O}_b}\frac{\partial {O}_c}{\partial \delta_c}\{b,p_b\}_D+\left(1-\frac{\partial K_b}{\partial {O}_b}\frac{\partial {O}_b}{\partial \delta_b}\right)\left(\frac{\partial {O}_c}{\partial c}\{b,c\}_D+\frac{\partial {O}_c}{\partial p_c}\{b,p_c\}_D\right)\bigg],
\end{align}
and
\begin{align}\label{dynpb}
&\{p_b,O_b\}_D=-\frac{1}{\beta}\bigg[\frac{\partial {O}_b}{\partial b}\left(1-\frac{\partial K_c}{\partial {O}_c}\frac{\partial {O}_c}{\partial \delta_c}\right)\{b,p_b\}_D+\frac{\partial K_b}{\partial {O}_c}\frac{\partial {O}_b}{\partial \delta_b}\left(\frac{\partial {O}_c}{\partial c}\{c,p_b\}_D-\frac{\partial {O}_c}{\partial p_c}\{p_b,p_c\}_D\right)\bigg], \nonumber\\& \{p_b,O_c\}_D=-\frac{1}{\beta}\bigg[\frac{\partial {O}_b}{\partial b}\frac{\partial K_c}{\partial {O}_b}\frac{\partial {O}_c}{\partial \delta_c}\{b,p_b\}_D+\left(1-\frac{\partial K_b}{\partial {O}_b}\frac{\partial {O}_b}{\partial \delta_b}\right)\left(\frac{\partial {O}_c}{\partial c}\{c,p_b\}_D-\frac{\partial {O}_c}{\partial p_c}\{p_b,p_c\}_D\right)\bigg],
\end{align}
where we have defined the function 
\begin{align}
\beta=\left(1-\frac{\partial K_b}{\partial {O}_b}\frac{\partial {O}_b}{\partial \delta_b}\right)\left(1-\frac{\partial K_c}{\partial {O}_c}\frac{\partial {O}_c}{\partial \delta_c}\right)-\frac{\partial K_b}{\partial {O}_c}\frac{\partial K_c}{\partial {O}_b}\frac{\partial {O}_b}{\partial \delta_b}\frac{\partial {O}_c}{\partial \delta_c},
\end{align}
which we assume to be non-zero (by suitably choosing $K_b$ and $K_c$, if necessary). Similar equations hold for the Dirac brackets between $c$, $p_c$, and the partial Hamiltonians, after interchanging the roles of $(b,p_b,\delta_b,{O}_b,K_b)$ by $(c,p_c,\delta_c,{O}_c,K_c)$. If one inserts the explicit expression of the Dirac brackets between connection and triad variables, and takes into account again that $P_{\delta_b}$ and $P_{\delta_c}$ Poisson commute with the partial Hamiltonians in the extended phase space, one finally obtains
\begin{align}\label{bpbdyn}
\{b,{O}_b\}_D=\gamma \frac{\partial {O}_b}{\partial p_b},\qquad \{p_b,{O}_b\}_D=-\gamma \frac{\partial {O}_b}{\partial b}, \qquad \{b,{O}_c\}_D=0, \qquad \{p_b,{O}_c\}_D=0, \\ \label{cpcdyn} \{c,{O}_b\}_D=0,\qquad \{p_c,{O}_b\}_D=0,\qquad  \{c,{O}_c\}_D=2\gamma \frac{\partial {O}_c}{\partial p_c},\qquad \{p_c,{O}_c\}_D=-2\gamma \frac{\partial {O}_c}{\partial c}.
\end{align}
Since these Dirac brackets define Hamiltonian flows on the reduced phase space, the above equations determine the reduced dynamics of the connection and triad variables. It is straightforward to realize that they provide precisely the equations of motion that one obtains from the Hamiltonian $H^{\text{eff}}_{\text{AOS}}$ in a Kantowski-Sachs type of cosmology if one treats the quantum parameters $\delta_b$ and $\delta_c$ as pure constants (to be later evaluated in terms of the constant of motion $m$ on each solution). In other words, the equations that we have obtained reproduce exactly the dynamics of the AOS model for the black hole interior. In this sense, a rigorous Hamiltonian derivation of these equations is possible because the change in the symplectic structure on $\bar{\Gamma}_{\text{ext}}$ with respect to $\Gamma_{\text{KS}}$ actually compensates the implicit dependence of the quantum parameters $\delta_b$ and $\delta_c$ on the partial Hamiltonians.

\section{Loop quantization}\label{sec3}

From our discussion, it seems evident that the deduction of the AOS equations from an effective Kantowski-Sachs Hamiltonian with LQC corrections [see Eq. \eqref{effectiveH}] requires that the symplectic algebra of the Ashtekar-Barbero variables be not canonical, unlike what happens in general relativity. In fact, depending on how one fixes the quantum parameters as functions of the partial Hamiltonians ${O}_b$ and ${O}_c$, this algebra can be very complicated and even display non-commutativity between different components of the connection and of the densitized triad, in spite of allowing a reduced dynamics of the AOS type. Finding a concrete quantum representation of such an algebra on a Hilbert space does not seem manageable using standard loop techniques. However, we know that the dynamics of the AOS model can be obtained from the effective Hamiltonian $H_\text{ext}^{\text{eff}}$ on the extended phase space $\Gamma_{\text{ext}}$, (at least) for an ample class of conditions fixing the gauge associated with the constraints $\Psi_b$ and $\Psi_c$. The connection and triad variables indeed form a canonical set in the subspace of $\Gamma_{\text{ext}}$ coordinatized by them. In this sense, a promising line of attack for the loop quantization of a black hole that leads to the AOS model in effective regimes is to consider on $\Gamma_{\text{ext}}$ the extended Hamiltonian
\begin{align}\label{exH}
H_{\text{ext}}=\underline{N}L_o\left(O^{\text{KS}}_b-O^{\text{KS}}_c\right)+\lambda_b \Psi^{\text{KS}}_b +\lambda_c \Psi^{\text{KS}}_c,
\end{align}
where $O^{\text{KS}}_b$ and $O^{\text{KS}}_c$ are the (densitized) partial Hamiltonians of the Kantowski-Sachs cosmologies in general relativity [see Eq. \eqref{KSHam}], and the constraints $\Psi^{\text{KS}}_b$ and $\Psi^{\text{KS}}_c$ are defined as
\begin{align}
\Psi^{\text{KS}}_b=K_b\left(O^{\text{KS}}_b,O^{\text{KS}}_c\right) - \delta_b, \qquad \Psi^{\text{KS}}_c=K_c\left(O^{\text{KS}}_b,O^{\text{KS}}_c\right) - \delta_c.
\end{align}
They incorporate the choice of quantum parameters that allows to reach the AOS model in the regime where these parameters have small absolute values [see Eqs. \eqref{deltas} and \eqref{deltasext}].  In the following, we will address the loop quantization of the canonical algebra of the extended phase space, and construct a quantum representation of the constraints contained in the Hamiltonian $H_{\text{ext}}$ following LQC techniques.

In LQG, one looks for a quantum representation of the Poisson algebra of holonomies of the Ashtekar-Barbero connection along edges and fluxes of the densitized triad through $2$-dimensional surfaces \cite{ashlewlqg,lqgThiemann}. In a geometry of the Kantowski-Sachs type [see Eq. \eqref{KSmetric}], if one focuses on surfaces delimited by edges in any of the coordinate directions $x$, $\theta$, and $\phi$, these fluxes are completely described by the variables $p_b$ and $p_c$. On the other hand, the matrix elements of holonomies of the connection along edges in the coordinate directions $\theta$ and $x$ are completely determined by complex exponentials of the form
\begin{align}
\mathcal{N}_{\mu_b}=e^{ib\mu_b/2},\qquad \mathcal{N}_{\mu_c}=e^{ic\mu_c/2}, \qquad \mu_b,\mu_c\in \mathbb{R}.
\end{align}
The holonomies along edges in the coordinate direction $\phi$, however, display a more complicated dependence on the connection variables $b$ and $c$, owing to the dependence of the component $A^{i}_{\phi}$ on the polar angle $\theta$ [see Eq. \eqref{AB}]. Nonetheless, we do not need them for the loop quantization of the system and, hence, we do not show their explicit expression here. All the relevant information about the Ashtekar-Barbero connection in the Kantowski-Sachs cosmologies is captured by the complex exponentials $\mathcal{N}_{\mu_b}$ and $\mathcal{N}_{\mu_c}$. These satisfy a Poisson algebra with the triad variables $p_b$ and $p_c$ which can be interpreted as two copies of the algebra used to describe homogeneous and isotropic systems in LQC \cite{abl}. Its triad representation is usually called {\sl polymeric representation}, and it can be defined on (two copies of) the Hilbert space of square integrable functions on the real line with respect to the discrete measure \cite{ze}. Specifically, calling this Hilbert space $\mathcal{H}^{\text{kin}}_{\text{LQC}}$ and denoting its basis elements as the kets $|\mu_b,\mu_c\rangle$, with $\mu_b,\mu_c\in \mathbb{R}$, the polymeric representation of the Poisson algebra of pairs $(\mathcal{N}_{\mu_b},p_b)$ and $(\mathcal{N}_{\mu_c},p_c)$ in Kantowski-Sachs is given by \cite{ABbh}
\begin{align}\label{polymeric}
\hat{\mathcal{N}}_{\mu'_b}|\mu_b,\mu_c\rangle=|\mu_b+\mu'_b,\mu_c\rangle,\qquad \hat{p}_b|\mu_b,\mu_c\rangle=\frac{\gamma \mu_b}{2}|\mu_b,\mu_c\rangle,\nonumber \\ \hat{\mathcal{N}}_{\mu'_c}|\mu_b,\mu_c\rangle=|\mu_b,\mu_c+\mu'_c\rangle,\qquad \hat{p}_c|\mu_b,\mu_c\rangle=\gamma \mu_c |\mu_b,\mu_c\rangle.
\end{align}
For the canonical pairs $(\delta_b,p_{\delta_b})$ and $(\delta_c,p_{\delta_c})$ that describe the rest of the extended phase space, we adopt a continuous Schrödinger representation, with Hilbert spaces simply given by $L^2(\mathbb{R},d\delta_b)$ and $L^2(\mathbb{R},d\delta_c)$. This choice completes the quantization of the canonical algebra on $\Gamma_{\text{ext}}$, paralleling the standard procedure followed in homogeneous and isotropic LQC minimally coupled to a scalar field. In total, the kinematic Hilbert space for the theory is given by $\mathcal{H}_T^{\text{kin}}=\mathcal{H}^{\text{kin}}_{\text{LQC}}\otimes L^2(\mathbb{R},d\delta_b)\otimes L^2(\mathbb{R},d\delta_c)$. Generalized basis elements of this space are denoted as $|\mu_b,\mu_c,\delta_b,\delta_c\rangle$ and they are normalized such that
\begin{align}
\langle \mu_b,\mu_c,\delta_b,\delta_c | \mu'_b,\mu'_c,\delta'_b,\delta'_c\rangle=\delta_{\mu_b,\mu'_b}\delta_{\mu_c,\mu'_c}\delta(\delta_b-\delta'_b)\delta(\delta_c-\delta'_c),
\end{align}
where $\delta_{y,y'}$ is the Kronecker delta and $\delta(y-y')$ is the Dirac delta. To study the quantum dynamics of the system, we need to find a representation of the constraints contained in the Hamiltonian $H_{\text{ext}}$ on the kinematic Hilbert space.

The representation of the holonomy-flux algebra displayed in Eq. \eqref{polymeric} is discrete, so one cannot define an operator representation for the connection variables $b$ and $c$. Considering that the partial Hamiltonians of the Kantowski-Sachs cosmologies depend on these variables through powers of them, their quantum representation as operators on the kinematic Hilbert space requires a regularization procedure. This situation is analogous to the one found in full LQG, where a well established regularization exists \cite{ashlewlqg,lqgThiemann,Thiemannreg}. Inspired by the main ideas driving the formulation of LQC, we follow this procedure by adapting it to the considered cosmological spacetimes. After imposing the cosmological symmetries, all the dependence of the Kantowski-Sachs Hamiltonian on the Ashtekar-Barbero connection $A^{i}_{\alpha}$ can be expressed in terms of its associated curvature tensor
\begin{align}
F^{i}_{\alpha\beta}=2\partial_{[\alpha} A^{i}_{\beta]}+\epsilon^{i} {}_{jk}A^{j}_{\alpha} A^{k}_{\beta},
\end{align}
or, rather, of an integrated version of it. If $\square(\alpha,\beta)$ is a coordinate rectangle in the plane $\alpha$--$\beta$, it is possible to relate the holonomy circuit along the edges that delimit it with the contribution from the corresponding curvature components in the limit in which the rectangle shrinks to a point (as much as it is physically allowed). This type of relations are traditionally employed in LQG to provide a regularization of the Hamiltonian \cite{Thiemannreg}. In Appendix \ref{appendix:Hreg} we explicitly show how to mimic such a regularization in the case of Kantowski-Sachs cosmologies, following the prescriptions used in the AOS model to fix a minimum physical area \cite{AOS2}. In short, the Hamiltonian is regularized by first postulating that there exists a certain minimum non-vanishing value of the physical area of $\square(\alpha,\beta)$, and then truncating the relations between the curvature components and the holonomy circuit at dominant order in the size of the edges of $\square(\alpha,\beta)$. In this process, the lengths of these edges are fixed in terms of the quantum parameters $\delta_b$ and $\delta_c$. As a result, it turns out that the regularized version of the Kantowski-Sachs Hamiltonian can be written precisely as the effective one $H^{\text{eff}}_{\text{AOS}}$, given in Eq. \eqref{effectiveH}. Since this function contains regularized versions ${O}_b$ and ${O}_c$ of the partial Hamiltonians $O^{\text{KS}}_c$ and $O^{\text{KS}}_b$, we directly prescribe that the regularization of the constraints $\Psi^{\text{KS}}_b$ and $\Psi^{\text{KS}}_c$ for their quantization is attained by simply replacing them with $\Psi_b$ and $\Psi_c$, given in Eq. \eqref{AOSconstraints}.

The regularized versions of the partial Hamiltonians depend on the connection variables $b$ and $c$ only through complex exponentials of the form $\mathcal{N}_{2\delta_b}$ and $\mathcal{N}_{2\delta_c}$. Therefore, we can construct operators for them using the basis of the kinematic Hilbert space with generalized elements $|\mu_b,\mu_c,\delta_b,\delta_c\rangle$ and the polymeric representation of the holonomy-flux algebra given in Eq. \eqref{polymeric}. For this purpose, we first define the quantum representation of the phase space functions $\sin(\delta_b b)$ and $\sin(\delta_c c)$ through their action on the basis elements:
\begin{align}
\widehat{\sin(\delta_b b)}  | \mu_b,\mu_c,\delta_b,\delta_c  \rangle=\frac{1}{2i}\left(  | \mu_b+2\delta_b,\mu_c,\delta_b,\delta_c \rangle - |\mu_b-2\delta_b,\mu_c,\delta_b,\delta_c\rangle\right), \nonumber \\
\widehat{\sin(\delta_c c)} |\mu_b,\mu_c,\delta_b,\delta_c\rangle=\frac{1}{2i}\left(|\mu_b,\mu_c+2\delta_c,\delta_b,\delta_c\rangle -|\mu_b,\mu_c-2\delta_c,\delta_b,\delta_c\rangle\right).
\end{align}
Next, inspired by factor ordering prescriptions that have proven useful in homogeneous and isotropic LQC (see Ref. \cite{mmo}, which employs the so-called MMO prescription), we introduce the operators
\begin{align}\label{Omegab}
\hat{\Omega}^{\delta_b}_b=\frac{1}{2\delta_b}|\hat{p}_b|^{1/2}\left[\widehat{\sin(\delta_b b)}\widehat{\text{sign}(p_b)}+\widehat{\text{sign}(p_b)}\widehat{\sin(\delta_b b)}\right]|\hat{p}_b|^{1/2},\\
\hat{\Omega}^{\delta_c}_c=\frac{1}{2\delta_c}|\hat{p}_c|^{1/2}\left[\widehat{\sin(\delta_c c)}\widehat{\text{sign}(p_c)}+\widehat{\text{sign}(p_c)}\widehat{\sin(\delta_c c)}\right]|\hat{p}_c|^{1/2},\label{Omegac}
\end{align}
for fixed non-zero $\delta_b$ and $\delta_c$\footnote{Since $\delta_b=0$ and $\delta_c=0$ are irrelevant points in the space of square integrable functions with respect to the Lebesgue measure, it is not necessary for any practical purpose to provide a detailed definition of the operators $\hat{\Omega}^{\delta_b}_b$ and $\hat{\Omega}^{\delta_c}_c$ for vanishing parameters $\delta_b$ and $\delta_c$.}, respectively, where the representation of functions of the triad variables $p_b$ and $p_c$ is defined in terms of the triad operators using the spectral theorem. We note that the operators $\hat{\Omega}^{\delta_b}_b$ and $\hat{\Omega}^{\delta_c}_c$ annihilate the kinematic states corresponding to $p_b=0$ and $p_c=0$, and completely decouple them from the rest of states (namely, from their complement in $\mathcal{H}_T^{\text{kin}}$). Moreover, these operators do not mix basis elements with positive and negative values of $\mu_b$ (and similarly for $\mu_c$). Then, without loss of generality, we will focus on states with $\mu_b>0$ and $\mu_c>0$ from now on. On each generalized eigenspace of the quantum parameters $\hat{\delta}_b$ and $\hat{\delta}_c$ with fixed non-zero eigenvalues $(\delta_b,\delta_c)$, the operators $\hat{\Omega}^{\delta_b}_b$ and $\hat{\Omega}^{\delta_c}_c$ only relate generalized states in our basis of the respective form
\begin{align}
|(\varepsilon_{b}+2n_b)|\delta_b|,\mu_c,\delta_b,\delta_c\rangle,\quad \varepsilon_{b}\in(0,2]\qquad \text{and}\qquad |\mu_b, (\varepsilon_{c} + 2n_c)|\delta_c|,\delta_b,\delta_c\rangle,\quad \varepsilon_{c}\in(0,2],
\end{align}
where $n_b,n_c\in\mathbb{N}$. Therefore, in each of those generalized eigenspaces, the operators preserve superselection sectors labelled by pairs $(\varepsilon_{b},\varepsilon_{c})$. These sectors are clearly separable. In the following, we will restrict our attention to any of them in each generalized eigenspace of the kinematic Hilbert space. Furthermore, let us assume for convenience that the operator $\hat{\Omega}^{\delta_b}_b$ admits an inverse on each superselection sector. Then, we can adopt the following representation of the partial Hamiltonians on the kinematic Hilbert space, determined by their action on basis elements:
\begin{align}
&\hat{O}_b|\mu_b,\mu_c,\delta_b,\delta_c\rangle=\hat{O}^{\delta_b}_b|\mu_b,\mu_c,\delta_b,\delta_c\rangle=-\frac{1}{2\gamma L_o}\bigg[\hat{\Omega}^{\delta_b}_b+\gamma^2 |\hat{p}_b|\left(\hat{\Omega}^{\delta_b}_b\right)^{-1}|\hat{p}_b|\bigg]|\mu_b,\mu_c,\delta_b,\delta_c\rangle,\label{Obdeltab} \\
&\hat{O}_c|\mu_b,\mu_c,\delta_b,\delta_c\rangle=\hat{O}^{\delta_c}_c|\mu_b,\mu_c,\delta_b,\delta_c\rangle=\frac{1}{\gamma L_o}\hat{\Omega}^{\delta_c}_c |\mu_b,\mu_c,\delta_b,\delta_c\rangle. \label{Ocdeltac}
\end{align}
Although other representations are admissible, the proposed representation is especially simple and keeps to a minimum the number of independent basic operators involved in the definition of the quantum constraints. Subtracting now the operators $\hat{O}_b$ and $\hat{O}_c$ (and multiplying the result by the fiducial length $L_o$), we obtain the Hamiltonian constraint operator for our extended Kantowski-Sachs model [see Eq. \eqref{exH}]. This constraint immediately decouples any state corresponding to vanishing triad variables. In this sense, looking at the expression of the metric given in Eq. \eqref{KSmetric}, one can say that the black hole singularity is already resolved at this level of the quantum theory (similarly to what happens in LQC with the Big Bang singularity). As for the remaining constraints $\Psi_b$ and $\Psi_c$, we can simply construct their quantum representation as operators by applying the spectral theorem to define the functions $K_b(\hat{O}_b,\hat{O}_c)$ and $K_c(\hat{O}_b,\hat{O}_c)$. For this definition (and in the following), we assume that $\hat{O}_b$ and $\hat{O}_c$ are self-adjoint, or that they can be replaced with suitable, unique self-adjoint extensions. This assumption is quite reasonable from a physical perspective if one recalls that, in the classical theory describing the Schwarzschild interior, the considered partial Hamiltonians determine the black hole mass.

\section{Physical states} \label{sec:phys}

The usual strategy to find physical states in LQC is to follow Dirac's approach for the quantization of constrained systems \cite{lqc2,lqc3,Dirac}. Given a kinematic Hilbert space and a representation of the constraints as operators on it, one focuses on the algebraic dual of a dense subset of this space, on which the constraints are defined by the adjoint action. Physical states are sought there by demanding that they be annihilated by the constraints. This set of physical states should eventually be provided with a Hilbert space structure. In what remains, we will follow this procedure to formally characterize the physical states of our quantization of the extended Kantowski-Sachs model.

In the search of physical states, it is most convenient to know the spectral properties of the constraint operators. For the purposes of this work, we will make the following assumptions (for more details about the spectral analysis of linear operators on a Hilbert space, see Refs. \cite{galindo,simon}):
\begin{itemize}
\item[i)] For all values of the parameters $\delta_b$ and $\delta_c$, the operators $\hat{O}^{\delta_b}_b$ and $\hat{O}^{\delta_c}_c$ are self-adjoint (or can be replaced with unique self-adjoint extensions). As a consequence, their spectra are real and only contain a continuous part and/or a point part. 
\item[ii)] The point parts of the spectra of $\hat{O}^{\delta_b}_b$ and $\hat{O}^{\delta_c}_c$ are discrete. This means that each eigenvalue is an isolated point of the real line and has finite multiplicity.
\item[iii)] The continuous parts of the spectra of $\hat{O}^{\delta_b}_b$ and $\hat{O}^{\delta_c}_c$ are absolutely continuous. In other words, they do not contain any singular part in which the spectral measure is not equivalent to the Lebesgue one. 
\end{itemize}
These three assumptions are quite reasonable and are actually shared by most of the Hamiltonian operators of physical quantum theories.

A remarkable property of the operator $\hat{\Omega}_c^{\delta_c}$ that will be extremely important in our quantum construction is that its spectrum turns out to be independent of the value of the parameter $\delta_c$ for any given superselection sector, characterized by a value of $\varepsilon_{c}\in (0,2]$. To prove this statement, let us consider the following change of the polymeric basis in each generalized eigenspace of the operators $\hat{\delta}_b$ and $\hat{\delta}_c$:
\begin{align}
|\mu_b,\mu_c,\delta_b,\delta_c\rangle \to |\tilde\mu_b,\tilde\mu_c,\delta_b,\delta_c\rangle, \qquad \tilde{\mu}_b=\frac{\mu_b}{|\delta_b|}, \qquad \tilde{\mu}_c=\frac{\mu_c}{|\delta_c|}.
\end{align}
Under the action of the triad operator $\hat{p}_c$, the new basis states get multiplied by $\gamma|\delta_c|\tilde{\mu}_c$. Thus, the action of $|\hat{p}_c|/|\delta_c|$, or of any (possibly fractional) power of it, its independent of the value of the parameter $\delta_c$. On the other hand, the operator $\widehat{\sin(|\delta_c| c)}$ simply generates constant 2-unit shifts in $\tilde{\mu}_c$ on the new basis elements, while the action of $\widehat{\text{sign}(p_c)}$ does not depend on $\delta_c$ either. Therefore, we conclude that the action of $\hat{\Omega}_c^{\delta_c}$, given by Eq. \eqref{Omegab}, is indeed independent of the (fixed) value of $\delta_c$ associated with the generalized eigenspace where it is defined. Hence, the same happens with its spectrum. Obviously, this applies as well to the spectrum of $\hat{O}_c^{\delta_c}$, because the two operators differ only by a multiplicative constant independent of the parameter. 

Clearly, the spectrum of $\hat{\Omega}_b^{\delta_b}$ is also independent of the value of $\delta_b$, since this operator is defined in a completely analogous way. However, this conclusion does not extend to the partial Hamiltonian $\hat O_b^{\delta_b}$. From the definition \eqref{Obdeltab} of this operator, we see that, up to irrelevant multiplicative constants, its action on a generalized state $|\tilde\mu_b,\tilde\mu_c,\delta_b,\delta_c\rangle$ differs from that of $\hat \Omega_b^{\delta_b}$ by the addition of another operator. This operator is proportional to the symmetrized product of the inverse of $\hat \Omega_b^{\delta_b}$ with powers of $|\hat{p}_b|/|\delta_b|$, all of them with actions that are independent of $\delta_b$ according to our discussion, but the proportionality factor is actually $\gamma^2\delta_b^2$. Given this dependence on the regularization parameter, it is reasonable to at least expect a smooth variation of the action of $\hat O_b^{\delta_b}$ with $\delta_b$. Moreover, we would expect that the term that contains the contribution of $\delta_b^2$ in $\hat O_b^{\delta_b}$ can be treated as a perturbation of $\hat{\Omega}_b^{\delta_b}$ when $|\delta_b|$ is small. 

Since the partial Hamiltonian operators are self-adjoint, we can employ the spectral theorem and decompose any state of the kinematic Hilbert space in terms of their (generalized) eigenfunctions \cite{galindo,simon}. Then, taking into account the spectral properties that we have assumed for these operators, we can write any dual element $(\psi|$ of the subset of states spanned by the (rescaled) basis elements $|\tilde\mu_b,\tilde\mu_c,\delta_b,\delta_c\rangle$ as
\begin{align}\label{psi}
(\psi|=\int_\mathbb{R}d\delta_b\int_\mathbb{R}d\delta_c\int_\mathbb{R}D\rho_{\delta_b}\int_\mathbb{R}Dm\sum_{\tilde\mu_b,\tilde\mu_c}\psi\left(\delta_b,\delta_c,\rho_{\delta_b},m\right)e^{\varepsilon_{b}}_{\rho_{\delta_b}}(\tilde\mu_b)\bar{e}^{\varepsilon_{c}}_{m}(\tilde\mu_c)\langle \tilde\mu_b,\tilde\mu_c,\delta_b,\delta_c|.
\end{align}
Let us explain this formula. First of all, we have restricted our attention to only one superselection sector of the rescaled polymeric basis, which is the same in all of the generalized subspaces of fixed $(\delta_b,\delta_c)$. This sector is labelled by the pair $(\varepsilon_b,\varepsilon_c)$. Allowing a dependence of these labels on the parameters $\delta_b$ and $\delta_c$ seems artificial if we are interested in studying the physical dynamics of the system. Indeed, it is easy to check that the action of the partial Hamiltonian operators, and therefore of all the constraint operators, leave these superselection sectors invariant. Moreover, a similar restriction has been adopted in many discussions in LQC (where a homogeneous scalar field plays a similar mathematical role as the parameters $\delta_b$ and $\delta_c$), assuming that the final results about measurable physical observables are not sensitive to the choice of a specific superselection subspace \cite{APS,mmo,mop}. The sums over $\tilde\mu_b$ and $\tilde\mu_c$ in Eq. \eqref{psi} run over each of the semilattices associated respectively with the labels $\varepsilon_{b}$ and $\varepsilon_{c}$. In addition, $D\rho_{\delta_b}$ and $Dm$ respectively denote the spectral measures associated with $\hat{O}^{\delta_b}_b$ and $\hat{O}^{\delta_c}_c$, whereas $\rho_{\delta_b}$ and $m$ stand for points of their spectra. Our notation emphasizes that the spectral measure of $\hat{O}^{\delta_c}_c$ is independent of the parameter $\delta_c$ and that the eigenvalues of this operator are directly related with the black hole mass. On the other hand, $e^{\varepsilon_{b}}_{\rho_{\delta_b}}(\tilde\mu_b)$ and $\bar{e}^{\varepsilon_{c}}_{m}(\tilde\mu_c)$ are the respective eigenfunctions of the operators $\hat{O}^{\delta_b}_b$ and $\hat{O}^{\delta_c}_c$ in the polymeric representation with our choice of basis elements. Unlike what occurs in the case of $\bar{e}^{\varepsilon_{c}}_{m}(\tilde\mu_c)$, we expect the eigenfunction $e^{\varepsilon_{b}}_{\rho_{\delta_b}}(\tilde\mu_b)$ to depend on the value of the parameter $\delta_b$, since the same happens with the action of the operator $\hat{O}^{\delta_b}_b$. For simplicity in the notation, we will not explicitly indicate this dependence. Finally, using the assumed properties of the partial Hamiltonian operators, we can always write the spectral measures $D\rho_{\delta_b}$ and $Dm$ directly as the discrete and the Lebesgue measures, respectively in the discrete and absolutely continuous parts of the spectrum. The change to these more familiar measures from the original spectral ones would be given by the so-called Radon-Nikodym derivative \cite{measure}, and it can be absorbed in the definition of the eigenfunctions $e^{\varepsilon_{b}}_{\rho_{\delta_b}}(\tilde\mu_b)$ and $\bar{e}^{\varepsilon_{c}}_{m}(\tilde\mu_c)$, convention that we adopt in the following.  

The simultaneous imposition of the three constraint operators of our system (namely $\hat{O}_b-\hat{O}_c$, $\hat{\Psi}_b$, and $\hat{\Psi}_c$) implies that the wave function $\psi$ must have the form
\begin{align}\label{phys1}
\psi\left(\delta_b,\delta_c,\rho_{\delta_b},m\right)=\xi (m)\delta_D\left[\delta_b-K_b(m,m)\right]\delta_D\left[\delta_c-K_c(m,m)\right]\delta_D(m-\rho_{\delta_b}),
\end{align}
in principle without restrictions on the function $\xi$. Here, $\delta_D$ denotes a general delta distribution, the discrete or continuous nature of which depends on the integration measure of the functional space where it is defined. In the following, we discuss the restrictions that these delta distributions impose on the physical states. 

Consider a fixed (but otherwise arbitrary) value of the parameter $\delta_b$, and let $\text{Sp}_b[\delta_b]$ and $\text{Sp}_c$ be the respective spectra of the operators $\hat{O}^{\delta_b}_b$ and $\hat{O}^{\delta_c}_c$. In principle, these spectra might even depend on the superselection semilattices chosen for the construction of the partial Hamiltonian operators. Recall that the spectrum of $\hat{O}^{\delta_c}_c$ is independent of $\delta_c$ and that we have fixed the corresponding superselection semilattice for all possible values of the pair $(\delta_b,\delta_c)$. The last delta in Eq. \eqref{phys1} then restricts the search for physical states to the intersection of the two considered spectra. According to our assumptions, this set will generally contain continuous intervals and a set of isolated points on the real line. The intervals can only arise from common continuous parts of $\text{Sp}_b[\delta_b]$ and $\text{Sp}_c$. On the other hand, the set of isolated points will have a contribution coming from the intersection of the discrete (continuous) part of $\text{Sp}_b[\delta_b]$ and the continuous (discrete) part of $\text{Sp}_c$. The rest of this set is simply the intersection of the discrete parts of $\text{Sp}_b[\delta_b]$ and $\text{Sp}_c$, which will only occur exceptionally. We call $\text{ISp}[\delta_b]$ the subset that is formed by this last intersection and by the commented continuous intervals. Since an isolated point in the continuous part of $\text{Sp}_b[\delta_b]$ or $\text{Sp}_c$ has zero measure, we limit our considerations to contributions in which the integration over the (generalized) eigenvalues $\rho_{\delta_b}$ and $m$ in Eq. \eqref{psi} is restricted to $\text{ISp}[\delta_b]$. Hence, the corresponding physical states of the model have their support on $\text{ISp}[\delta_b]$. 

We now turn our attention to the remaining deltas in Eq. \eqref{phys1}, involving the regularization parameters. For this, we define the pair of functions
\begin{align}
\tilde{K}_b(m)=K_b(m,m), \qquad \tilde{K}_c(m)=K_c(m,m),\qquad m\in \mathbb{R}.
\end{align}
Concerning the second delta distribution in that equation, we note that, as long as the domain of $\tilde{K}_c$ contains $\text{ISp}[\delta_b]$, something that can be guaranteed by construction of this function, there always exists some $\delta_c\in\mathbb{R}$ that coincides with the resulting value of $\tilde{K}_c(m)$. However, the value of $\tilde{K}_b$, when applied to an arbitrary point of $\text{ISp}[\delta_b]$, need not be equal to $\delta_b$, as it is required by the first delta distribution in Eq. \eqref{phys1}. Taking into account this imposition and considering all possible values of $\delta_b$, it follows that the support of the physical states is restricted to the subset of $\text{Sp}_c$ given by 
\begin{align}
\text{CSp}=\left\{ m \in \text{ISp}\big[\delta_b=\tilde{K}_b(m)\big] \right\}.
\end{align}
In other words, $m$ belongs to $\text{CSp}$ if it belongs to the intersection of the (absolutely) continuous parts of $\text{Sp}_c$ and $\text{Sp}_b[\tilde{K}_b(m)]$ or to the intersection of the discrete parts of these two sets. In total, we conclude that the physical states should admit the following formal expression:
\begin{align}\label{phys2}
(\psi|=\int_{\text{CSp}} Dm \, \xi (m)\sum_{\tilde\mu_b,\tilde\mu_c} e^{\varepsilon_{b}}_{m}(\tilde\mu_b)\big |_{\delta_b=\tilde{K}_b(m)}\, \bar{e}^{\varepsilon_{c}}_{m}(\tilde\mu_c) \langle \tilde\mu_b,\tilde\mu_c,\delta_b=\tilde{K}_b(m),\delta_c=\tilde{K}_c(m)|.
\end{align}

The physical states of the system are then completely characterized by wave functions $\xi$ with support on the spectral set $\text{CSp}$. Therefore, we see that physical wave profiles of the quantum AOS model are simply functions of the black hole mass. The allowed values for this mass must belong to a very specific subset of the real line determined by the spectral properties of the loop quantum operators. The measure $Dm$ is a natural candidate to endow this set of physical states with a Hilbert structure, namely that of square integrable functions $\xi$ over the black hole mass. Physical observables will be represented by linear operators on this Hilbert space. Obvious examples are the multiplicative operator representing the black hole mass and translation operators between points of $\text{CSp}$.

We end this section commenting on the viability that $\text{CSp}$ contains points of arbitrarily large value that belong to sets $\text{ISp}[\delta_b]$ with small $|\delta_b|$. This part of the support of the physical states corresponds to large black hole masses that are mapped to small quantum parameters. It is evident that the existence of such a region is necessary for the quantum theory to be physically satisfactory. First of all, let us recall that, for large absolute values of its argument, the function $\tilde{K}_b$ follows the behavior given in the first identity of Eq. \eqref{deltasext}. Thus, for very large $|m|$, it is a monotonically decreasing function and the mapping from eigenvalues to quantum parameters $\delta_b$ is one-to-one (in fact, a similar statement is valid for $\tilde{K}_c$ and $\delta_c$). Furthermore, the mapping clearly brings, in absolute value, large eigenvalues (masses) to small quantum parameters. It follows that the physical states of the theory can only describe realistic black holes, with masses that take any arbitrarily large value and are mapped to small quantum parameters, if the union of the sets $\text{ISp}[\delta_b]$ for small values of $|\delta_b|$ contains a continuous interval that is unbounded from above in absolute value. This condition, in turn, translates into an important restriction on the admissible spectral properties of the operators $\hat{O}^{\delta_c}_c$ and $\hat{O}^{\delta_b}_b$. The spectrum of the former must contain an absolutely continuous part that is unbounded from above in absolute value, while the spectrum of the latter must contain as well an absolutely continuous part for small values of $|\delta_b|$ such that it becomes unbounded from above (in absolute value) when this parameter is negligible. This result simply follows from the fact that, according to our previous discussion, the union of the sets $\text{ISp}[\delta_b]$ over any interval of the parameter $\delta_b$ may only contain isolated points coming from the intersection of the discrete spectra of the partial Hamiltonian operators, or intervals coming from the intersection of their continuous spectra. Thus, we see that only partial Hamiltonian operators with a very specific type of spectra lead to a quantum theory that can describe physically realistic black holes. In this context, it is worth mentioning that, if a spectral analysis of these operators is carried out in the future, one can rule out or put forward their definition as physically acceptable in a straightforward manner using our results. In fact, with this procedure, one can even test the very validity of the loop quantum representation of the model, if factor ordering ambiguities turn out to be irrelevant for the spectral properties of the partial Hamiltonians.

\section{Conclusions}\label{conclu}

In this work we have provided a consistent Hamiltonian formulation of the AOS model for the effective loop quantum extension of a non-rotating and non-charged black hole interior. Under such a Hamiltonian framework, we have set the main steps for the quantization of this type of spacetimes following LQC techniques, including definitions of the constraint operators and a preliminary study of the physical states. Our work finds an answer to some caveats posed to the AOS model since it was first proposed, about the consistency of its dynamics from a canonical perspective: if the effective Hamiltonian includes a dependence on constants of motion of the system, through loop quantum regularization parameters, their dependence on phase space should be taken into account when deriving the equations of motion. We show that, in order to consistently derive this dynamics from a Hamiltonian formulation, one either has to give up the symplectic structure of the black hole interior in general relativity, or extend its phase space to include the quantum parameters introduced to regularize the Hamiltonian according to LQG principles. Namely, there is no way of reproducing exactly the purely relativistic formulation of the system with a gauge fixing and the subsequent reduction of this extended phase space. Since the relativistic algebra of connections and triads is the fundamental basis for the formulation of LQG in terms of holonomies and fluxes, we have decided to address the loop quantization of the extended system, which includes extra constraints, in addition to the relativistic one, codifying the relation between the quantum parameters and constants of motion.

Employing well developed prescriptions in LQC, we have promoted the partial Hamiltonian of the angular sector [i.e., the sector corresponding to $(c,p_c)$] to an operator in a polymeric representation. Remarkably, taking the counterpart of this operator for the radial sector as the only basic ingredient, we have been able to construct as well a representation for the other partial Hamiltonian, and therefore for the Hamiltonian constraint of the system (given by the difference between the operators of the two sectors). Moreover, using these partial Hamiltonian operators we have been able to represent almost straightforwardly the remaining constraints that fix the parameters $\delta_b$ and $\delta_c$. This quantization of the model has allowed us to formally deduce the general expression of the physical states, provided that the partial Hamiltonian operators have certain reasonable spectral properties. Specifically, we have assumed that they are self-adjoint (or can be replaced by a unique self-adjoint extension) and that their spectra only contain discrete and absolutely continuous parts. Based on similar constructions used in LQC, in all the considered contributions to the physical states we have adopted the same (rescaled) superselection subspaces. With this restriction, the spectrum of the partial Hamiltonian $\hat{O}_c^{\delta_c}$ turns out to be completely independent of the parameter $\delta_c$. Then, physical states can be completely characterized by a wave function of the black hole mass, with support on a very specific subset of the spectrum of the partial Hamiltonian of the radial sector. This subset is essentially the part of the intersection with the spectrum of the other partial Hamiltonian which is consistently mapped to the values of the parameter $\delta_b$. We have also discussed under which circumstances this subset includes the sector of interest with large masses. The most important condition is that the spectra of the operators that represent the partial Hamiltonians must include an interval extending to infinity, both in the case of the angular sector that is parameter independent and for negligibly small values of $\delta_b$ in the case of the radial Hamiltonian. Proving that this is indeed so would require an spectral analysis of our operators, something that will be carried out in future research. The verification of this condition, together with a suitable choice of the functions that relate the mass with the parameters (e.g. of the type chosen in the AOS model), would guarantee that there exist physical states corresponding to geometries with large black hole masses and, predictably, small local quantum effects. 

Our investigations constitute an important step towards the construction of a quantum theory for the interior of a non-rotating uncharged black hole, even despite the possible criticisms existing in the literature about the physical properties of the AOS model as an effective geometry. Such a quantum theory, in turn, is necessary in order to study how quantum gravitational phenomena may affect the behavior of possible perturbations around the black hole geometry. These perturbations are of great physical importance since they can be used, e.g., to describe the ringdown part of the gravitational wave signal of a non-rotating black hole merger. In the context of the AOS model, there already exists a number of works that address how the loop quantum corrections to the spacetime metric affect the gravitational perturbations and their quasinormal modes [see e.g. \cite{Mariamquasi,Greenquasi,Javiquasi}]. However, all those studies assume that the dynamics of these perturbations are ruled by their relativistic equation, with a background metric that is not exactly the Schwarzschild one. In order to really capture how the gravitational waves are affected by quantum gravity phenomena, it is necessary to quantize their contribution to the Hamiltonian of the system, together with the black hole background. This may well lead to dynamical equations for the perturbations that differ from the relativistic ones, as it has been shown to be the case in cosmology \cite{massmax}. Our work provides a first building block for this ambitious program, which requires a satisfactory quantum description of the black hole interior.

\acknowledgments

This work was partially supported by Project No. MICINN PID2020-118159GB-C41 from Spain. B.E.N. acknowledges financial support from the Standard Program of JSPS Postdoctoral Fellowships for Research in Japan. A. G.-Q. acknowledges that the project that gave rise to these results received the support of a fellowship from ``La Caixa'' Foundation (ID 100010434). The fellowship code is LCF/BQ/DR19/11740028.

\appendix

\section{Explicit expressions of the momenta on the extended phase space} \label{appendix:Pdeltas}

In the discussion of the gauge fixing that leads to the reduced phase space $\bar \Gamma_{\rm ext}$, we have defined a new set of canonical variables that facilitates the comparison with Ref.  \cite{AOS2}. Instead of the original connection-triad canonical pairs, it is convenient to regard the partial Hamiltonians $O_b$ and $O_c$ as the new configuration variables, introducing canonically conjugate momenta given by
\begin{align}
P_b&=- \frac{2L_o}{b_o}\tanh^{-1}\left[\frac{\cos( \delta_b b)}{b_o}\right]  -2L_o\ln \dfrac{\gamma |\delta_b|}{2} ,\\
P_c&=-\frac{L_o}{2} \ln\bigg|\dfrac{2p_c\sin(\delta_c c)}{\gamma L_o^2 \delta_c^2}\tan\left(\frac{\delta_c c}{2}\right)\bigg|,
\end{align}
where $b_o=\sqrt{1+\gamma^2 \delta_b^2}$. It is straightforward to verify that $O_b$, $O_c$, $P_b$, and $P_c$ form a canonical set with non-zero Poisson brackets equal to $\{O_b,P_b\}=1$ and $\{O_c,P_c\}=1$. This change of variables can be completed into a canonical transformation in the extended phase space $\Gamma_{\rm ext}$ by introducing new momenta $P_{\delta_b}$ and $P_{\delta_c}$ that not only are canonically conjugate to $\delta_b$ and $\delta_c$, respectively, but also Poisson commute with the two partial Hamiltonians and their associated momenta. The following momenta meet these conditions:
\begin{align}
P_{\delta_b}&=p_{\delta_b}- \frac{p_b}{\gamma \delta_b}\left[b-\dfrac{\sin(\delta_b b)}{\delta_b}-\dfrac{\gamma^2 \delta_b}{\sin(\delta_b b)}\right]  \nonumber\\
&-\frac{\gamma p_b}{b_o^2\sin(\delta_b b)}\left\{\cos(\delta_b b)+ \frac{1}{b_o}\tanh^{-1}\left[\frac{\cos(\delta_b b)}{b_o}\right]  \left[b_o^2-\cos^2(\delta_b b)\right]\right\},\\
P_{\delta_c}&=p_{\delta_c}-\dfrac{p_c}{2\gamma \delta_c }\left[c-\dfrac{\sin( \delta_c c)}{\delta_c}\right].
\end{align}

\section{Regularization of the Kantowski-Sachs Hamiltonian}\label{appendix:Hreg}

In this appendix we show how a systematic procedure following LQG techniques \cite{Thiemannreg}, and supplemented with the prescription of the AOS model to fix the minimum physical area, leads to the effective formula \eqref{effectiveH} as a regularization of the Kantowski-Sachs Hamiltonian. The Hamiltonian constraint of general relativity can be generally written in terms of the Ashtekar-Barbero variables as 
\begin{align}
\mathcal{C}=\frac{1}{16\pi \sqrt{h}}\left[\epsilon^{ij}{}_k F^{k}_{\alpha\beta}-2(1+\gamma^2)K^{i}_{[\alpha}K^{j}_{\beta]}\right]E^{\alpha}_i E^{\beta}_j,
\end{align}
where $h$ is the determinant of the spatial metric, $\epsilon^{ij}{}_k$ is the Levi-Civita symbol, and $K^{i}_{\alpha}$ is the triadic form of the extrinsic curvature of the spatial sections. In the case of Kantowski-Sachs cosmologies, it holds that $h=p_b^2 |p_c| \sin^2 \theta /L_o^{2}$. Moreover, imposing the symmetries of this type of spacetimes we get the following identity:
\begin{align}
2K^{i}_{[\alpha}K^{j}_{\beta]}E^{\alpha}_{i}E^{\beta}_{j}=\frac{1}{\gamma^{2}}\epsilon^{ij}{}_k F_{\alpha\beta}^k E^{\alpha}_i E^{\beta}_j+\frac{p_b^2}{\gamma^{2}L_o^{2}} \sin^2 \theta.
\end{align}
Taking this into account and recalling the form of the components of the densitized triad for Kantowski-Sachs, given in Eq. \eqref{KStriad}, we obtain a Hamiltonian constraint of the form
\begin{align}
\mathcal{C}=\frac{1}{8\pi \gamma^2 \sqrt{|p_c|}}\bigg[p_c \left(F^{1}_{x\theta}\sin\theta+F^{2}_{x\phi}\right)-\frac{p_b}{L_o}\left(\sin\theta+\gamma^2\sin\theta-F^{3}_{\theta\phi}\right)\bigg].
\end{align}

The above expression inside the square brackets is what we have to regularize adapting LQG techniques to our cosmological system. For this purpose, we first consider two holonomy circuits $h_{\square(x,\theta)}$ and $h_{\square(x,\phi)}$ along coordinate rectangles $\square(x,\theta)$ and $\square(x,\phi)$ in the directions $(x,\theta)$ and $(x,\phi)$ that enclose the respective regions $(0,L_o\bar\mu_x)\times(0,\pi\bar\mu_\theta)$ and $(0,L_o\bar\mu_x)\times(0,2\pi\bar\mu_\phi)$. The positive parameters $\bar\mu_x,\bar\mu_\theta$, and $\bar\mu_\phi$ will be related to the minimum physical area allowed by the AOS model. Taking into account the definition of holonomy along an arbitrary edge $e$
\begin{align}\label{holonomy}
h_{e}=\mathcal{P}\exp\int_{e}dx^{\alpha}A^{i}_{\alpha}\tau_{i},
\end{align}
where $\mathcal{P}$ denotes path ordering and $\tau_{i}$ are the generators of the Lie algebra $su(2)$, it is possible to show that
\begin{align}\label{regeasy}
F^{i}_{x\theta}=-2\lim_{\bar\mu_x,\bar\mu_\theta \rightarrow 0}\frac{\text{Tr}\left[h_{\square(x,\theta)}\tau^{i}\right]}{\pi L_o \bar\mu_x\bar\mu_\theta},\qquad F^{i}_{x\phi}=-2\lim_{\bar\mu_x,\bar\mu_\phi \rightarrow 0}\frac{\text{Tr}\left[h_{\square(x,\phi)}\tau^{i}\right]}{2\pi L_o \bar\mu_x\bar\mu_\phi}.
\end{align}
Explicitly, we have that
\begin{align}\label{traceseasy}
2\text{Tr}\left[h_{\square(x,\theta)}\tau^{i}\right]=\pi\delta^i_1 \sin(\bar\mu_x c)\sin(\bar\mu_\theta b)+\mathcal{O}(3),\qquad 2\text{Tr}\left[h_{\square(x,\phi)}\tau^{i}\right]=2\pi\delta^i_2  \sin(\bar\mu_x c)\sin(\bar\mu_\phi b)\sin\theta+\mathcal{O}(3),
\end{align}
where $\mathcal{O}(n)$ contains any $n$th order product between the parameters $\bar\mu_x$, $\bar\mu_\theta$, and $\bar\mu_\phi$ or smaller terms, in the limit in which these parameters are negligibly small. Drawing direct inspiration from LQC \cite{lqc2,lqc3}, we regularize the contributions from $F^{i}_{x\theta}$ and $F^{i}_{x\phi}$ in the Hamiltonian constraint by removing the limit of vanishing area in Eq. \eqref{regeasy} (thus prescribing a minimum non-zero physical area in the theory), and replacing the traces of the right-hand side by the dominant order terms appearing in Eq. \eqref{traceseasy}. 

The regularization procedure for the contribution of the curvature components $F^i_{\theta\phi}$ to the Hamiltonian constraint is a bit more involved than in standard LQC. This is due to the complexity of the holonomy along the $\phi$ direction as a function of $b$ and $\theta$. In particular, rather than the simple identities that hold for the other components of the curvature [see Eq. \eqref{regeasy}], in this case we use that
\begin{align}
\int_{\square_{n,\tilde{n}}(\theta,\phi)}d\theta d\phi \,F^{i}_{\theta\phi}=-2\text{Tr}\left[h_{\square_{n,\tilde{n}}(\theta,\phi)}\tau^{i}\right]
\end{align}
at dominant order in the limit of small parameters $\bar\mu_\theta$ and $\bar\mu_\phi$, where $\square_{n,\tilde{n}}(\theta,\phi)$ is the coordinate rectangle enclosing the region $(n\pi\bar\mu_\theta,n\pi\bar\mu_\theta+\pi\bar\mu_\theta)\times(2\tilde{n}\pi\bar\mu_\phi,2\tilde{n}\pi\bar\mu_\phi+2\pi\bar\mu_\phi)$, for any $n,\tilde{n}\in \mathbb{N}$. Explicitly, we have
\begin{align}
2\text{Tr}\left[h_{\square_{n,\tilde{n}}(\theta,\phi)}\tau^{i}\right]=\pi^3(1+2n) \delta_{3}^i\left[\bar\mu_\theta^2\bar\mu_\phi +\sin^2(\bar\mu_\theta b)\bar\mu_\phi+2\sin(\bar\mu_\theta b)\sin(\bar\mu_\phi b)\bar\mu_\theta\right]+\mathcal{O}(4).
\end{align}
Our prescription for the regularization of the contributions coming from $F^i_{\theta\phi}$ in the Hamiltonian constraint is then the following. Upon integrating $\mathcal{C}$ over the unit two-sphere coordinatized by the pair $(\theta,\phi)$, we cover it with as many coordinate rectangles of the form $\square_{n,\tilde{n}}(\theta,\phi)$ as possible and then make the replacement
\begin{align}
-\int_0^\pi d\theta \int_0^{2\pi}d\phi F^i_{\theta\phi}\longrightarrow \pi^3 \delta_{3}^i \, M\left[\frac{1}{\bar\mu_\phi}\right]\sum_{n=0}^{M[1/\bar\mu_\theta]-1}(1+2n)\left[\bar\mu_\theta^2\bar\mu_\phi +\sin^2(\bar\mu_\theta b)\bar\mu_\phi+2\sin(\bar\mu_\theta b)\sin(\bar\mu_\phi b)\bar\mu_\theta\right],
\end{align}
where $M[\cdot]$ denotes the natural number immediately greater than or equal to its argument.

The parameters $\bar\mu_x$, $\bar\mu_\theta$, and $\bar\mu_\phi$ play a key role in the two regularization procedures that we have introduced. Indeed, they define the minimum coordinate areas allowed in the system, so that the dynamical contribution of the Ashtekar-Barbero connection can be quantized in a discrete way. In this sense, from a fundamental perspective it is important that one makes a connection between the values of the parameters and the theory that motivates them, in this case LQG. In the AOS model, this is done by relating them in a very specific way with the minimum area $\Delta$ allowed by the spectra of the area operator in LQG. Concretely, on each solution of this effective black hole model, it is required that \cite{AOS2}\footnote{In Ref. \cite{AOS2} the authors implicitly assume that the parameters $\delta_b$ and $\delta_c$ are positive, whereas in our work we let them be real.}
\begin{align}\label{AOSdelta}
\Delta=2\pi |\delta_c| |\delta_b| |p_b|_{{}_{\mathcal{T}}},\qquad \Delta=4\pi \delta_b^2 |p_c|_{{}_{\mathcal{T}}},
\end{align}
where the subscript $\mathcal{T}$ indicates evaluation on the smooth transition surface that replaces the singularity, and the right-hand sides are {\emph {by definition}} the physical areas on this surface of, respectively, the  coordinate rectangles $\square(x,\phi)|_{\theta=\pi/2}$ and $\square(\theta,\phi)$ covering the coordinate regions $(0,L_o\bar\mu_x)\times(0,2\pi\bar\mu_\phi)$ and $(0,\pi\bar\mu_\theta)\times(0,2\pi\bar\mu_\phi)$. Since we want to construct a quantum theory that eventually leads to the AOS model in effective regimes, we naturally adapt this prescription to our model by demanding that $\bar\mu_x$, $\bar\mu_\theta$, and $\bar\mu_\phi$ are fixed in terms of the quantum parameters $\delta_b$ and $\delta_c$ in such a way that the physical areas of the considered rectangles equal the right-hand side of Eq. \eqref{AOSdelta}, evaluated off-shell. It is then the set of constraints $\Psi_b$ and $\Psi_c$ in our extended phase space formulation what relates these areas with the minimum value $\Delta$ allowed by LQG. A direct application of this prescription shows that
\begin{align}
|\delta_c|=\bar\mu_x,\qquad |\delta_b|=\bar\mu_\phi=\frac{\pi^2\bar\mu_\theta^2}{4}.
\end{align}

If we integrate the Hamiltonian density $N\mathcal{C}$ over the spatial hypersurfaces with topology $I\times S^2$ of the Kantowski-Sachs cosmologies and apply the regularization prescriptions described above, we finally obtain the following expression for the regularized Hamiltonian:
\begin{align}
H^{\text{reg}}_{\text{KS}}[N]=-N\frac{\sin(\delta_b b)}{\gamma\delta_b\sqrt{|p_c|}}\left[H^{\text{eff}}_{\text{AOS}}+\mathcal{O}(\delta_b)\right],
\end{align}
where $\mathcal{O}(\delta_b)$ denotes terms that are of the order of $\delta_b$ or smaller in the limit in which this parameter is small in absolute value. Taking into account that these subdominant terms vanish when the corrections coming from the loop quantization procedure are ignored, we simply remove them from our regularized expression of the Hamiltonian. Then, up to a global factor that can be absorbed in a suitable choice of densitization, we see that our regularization of the Kantowski-Sachs Hamiltonian using LQG techniques and adhering to the AOS prescription for the minimum non-zero physical area indeed is equivalent to the effective Hamiltonian of the AOS model.

It is worth mentioning that many replacements performed in our regularization amount to fixing ambiguities that are more involved than those arising in the case of homogeneous and isotropic cosmologies \cite{lqc2,lqc3}. Namely, the regularization chooses to keep some subdominant terms while neglecting others in the expansion of the traces of holonomy circuits over small areas, and also in the expansion of the resulting Hamiltonian in the limit of small quantum parameters $\delta_c$ and $\delta_b$. Our guideline for this choice has been to maintain the result as simple as possible with an eye on its polymeric quantization. To the best of our knowledge, this is the first explicit derivation of the Hamiltonian of the AOS model as a loop quantum regularization of the Kantowski-Sachs one, using closed holonomy circuits.


\begin{thebibliography}{299}

\bibitem{ashlewlqg}  A. Ashtekar and J. Lewandowski, Background independent quantum gravity: A status report, Class. Quantum Grav. {\bf{21}}, R53 (2004).

\bibitem{lqgThiemann} T. Thiemann, Modern Canonical Quantum General Relativity (Cambridge University Press, Cambridge, UK, 2007). 

\bibitem{abl} A. Ashtekar, M. Bojowald, and J. Lewandowski, Mathematical structure of loop quantum cosmology, Adv. Theor. Math. Phys. {\bf 7}, 233 (2003).

\bibitem{lqc3} G.A. Mena Marug\'an, A brief introduction to loop quantum cosmology, AIP Conf. Proc. {\bf 1130}, 89 (2009).

\bibitem{lqc2} A. Ashtekar and P. Singh, Loop quantum cosmology: A status report, Class. Quantum Grav. {\bf 28}, 213001 (2011).

\bibitem{hyb-pert1} M. Fern\'andez-M\'endez, G.A. Mena Marug\'an, and J. Olmedo, Hybrid quantization of an inflationary universe, Phys. Rev. D {\bf 86}, 024003  (2012).

\bibitem{AAN1}  I. Agullo, A. Ashtekar, and W. Nelson, Extension of the quantum theory of cosmological perturbations to the Planck era, Phys. Rev. D {\bf 87}, 043507 (2013).

\bibitem{AAN2}  I. Agullo, A. Ashtekar, and W. Nelson, The pre-inflationary dynamics of loop quantum cosmology: Confronting quantum gravity with observations, Class. Quantum Grav.  {\bf 30}, 085014 (2013).

\bibitem{Bojo2} A. Barrau, M. Bojowald, G. Calcagni, J. Grain, and M. Kagan, Anomaly-free cosmological perturbations in effective canonical quantum gravity, JCAP {\bf 05}, 051 (2015).

\bibitem{hyb-pert4} L. Castell\'o Gomar, M. Mart\'{\i}n-Benito, and G.A. Mena Marug\'an, Gauge-invariant perturbations in hybrid quantum cosmology, JCAP {\bf 06}, 045 (2015).

\bibitem{Hybridreview} B. Elizaga Navascu\'es and G.A. Mena Marug\'an,  Hybrid loop quantum cosmology: An overview, Front. Astron. Space Sci. {\bf 8}, 624824 (2021).

\bibitem{Bojo1} M. Bojowald, G. Calcagni, and S. Tsujikawa, Observational constraints on loop quantum cosmology,  Phys. Rev. Lett. {\bf 107}, 211302 (2011).

\bibitem{Ivan} I. Agullo and N.A. Morris, Detailed analysis of the predictions of loop quantum cosmology for the primordial power spectra, Phys. Rev. D {\bf 92}, 124040 (2015).

\bibitem{Edward} E. Wilson-Ewing, Testing loop quantum cosmology, C. R. Phys. {\bf 18}, 207 (2017).

\bibitem{hybpred1} L. Castell\'o Gomar, G.A. Mena Marug\'an, D. Mart\'{\i}n de Blas, and J. Olmedo, Hybrid loop quantum cosmology and predictions for the Cosmic Microwave Background, Phys. Rev. D {\bf 96}, 103528 (2017).

\bibitem{hybpred2} B. Elizaga Navascu\'es, D. Mart\'{\i}n de Blas, and G.A. Mena Marug\'an, The vacuum state of primordial fluctuations in hybrid loop quantum cosmology, Universe {\bf 4}, 98 (2018).

\bibitem{AshNe} A. Ashtekar, B. Gupt, D. Jeong, and V. Sreenath, Alleviating the tension in CMB using Planck-scale physics, Phys. Rev. Lett. {\bf 125}, 051302 (2020).

\bibitem{hybothers} B.-F. Li, J. Olmedo, P. Singh, and A. Wang, Primordial scalar power spectrum from the hybrid approach in loop cosmologies, Phys. Rev. D {\bf  102}, 126025 (2020).

\bibitem{JCAP} B. Elizaga Navascu\'es and G.A. Mena Marug\'an, Analytical investigation of pre-inflationary effects in the primordial power spectrum: From General Relativity to hybrid loop quantum cosmology, JCAP {\bf 09}, 030 (2021).

\bibitem{ABbh} A. Ashtekar and M. Bojowald, Quantum geometry and the Schwarzschild singularity, Class. Quantum Grav. {\bf 23}, 391 (2006).

\bibitem{Modestobh} L. Modesto, Loop quantum black hole, Class. Quantum Grav. {\bf 23} 5587 (2006).

\bibitem{CKbh} D. Cartin and G. Khanna, Wave functions for the Schwarzschild black hole interior, Phys. Rev. D {\bf 73} 104009 (2006).

\bibitem{BVbh} C.G. Boehmer and K. Vandersloot, Loop quantum dynamics of Schwarzschild interior, Phys. Rev. D {\bf 76}, 1004030 (2007).

\bibitem{Chioubh} D.W. Chiou, Phenomenological loop quantum geometry of the Schwarzschild black hole, Phys. Rev. D {\bf 78}, 064040 (2008).

\bibitem{CGPbh1} M. Campiglia, R. Gambini, and J. Pullin, Loop quantization of spherically symmetric midisuperspaces: The interior problem, AIP Conf. Proc. {\bf 977}, 52 (2008).

\bibitem{SKbh} S. Sabharwal and G. Khanna, Numerical solutions to lattice-refined models in loop quantum cosmology, Class. Quantum Grav. {\bf 25}, 085009 (2008).

\bibitem{BKdBbh} J. Brannlund, S. Kloster, and A. De Benedictis, The evolution of black holes in the mini-superspace approximation of loop quantum gravity, Phys. Rev. D {\bf 79}, 084023 (2009).

\bibitem{GOPbh} R. Gambini, J. Omedo, and J. Pullin, Quantum black holes in loop quantum gravity, Class. Quantum Grav. {\bf 31}, 095009 (2014).

\bibitem{DJSbh} N. Dadhich, A. Joe, and P. Singh, Emergence of the product of constant curvature spaces in loop quantum cosmology, Class. Quantum Grav. {\bf 32}, 185006 (2015).

\bibitem{HRbh} H.M. Haggard and C. Rovelli, Quantum-gravity effects outside the horizon spark black to white hole tunnelling, Phys. Rev. D {\bf 92}, 104020 (2015).

\bibitem{CGOPbh} M. Campiglia, R. Gambini, J. Olmedo, and J. Pullin, Quantum self-gravitating collapsing matter in a quantum geometry, Class. Quantum Grav. {\bf 33}, 18LT01 (2016).

\bibitem{CSbh} A. Corichi and P. Singh, Loop quantization of the Schwarzschild interior revisited, Class. Quantum Grav. {\bf 33}, 055006 (2016).

\bibitem{OSSbh} J. Olmedo, S. Saini, and P. Singh, From black holes to white holes: A quantum gravitational, symmetric bounce, Class. Quantum Grav. {\bf 34}, 225011 (2017).

\bibitem{Jerobh} J. Cortez, W. Cuervo, H.A. Morales-Técotl, and J.C. Ruelas, On effective loop quantum geometry of Schwarzschild interior, Phys. Rev. D {\bf 95}, 064041 (2017).

\bibitem{YKSbh} A. Yonika, G. Khanna, and P. Singh, Von-Neumann stability and singularity resolution in loop quantized Schwarzschild black hole, Class. Quantum Grav. {\bf 35}, 045007 (2018).

\bibitem{BCDHRbh} E. Bianchi, M. Christodoulou, F. D’Ambrosio, H.M. Haggard, and C. Rovelli, White holes as remnants: A surprising scenario for the end of a black hole, Class. Quantum Grav. {\bf 35}, 225003 (2018).

\bibitem{ABP} E. Alesci, S. Bahrami, and D. Pranzetti, Quantum gravity predictions for black hole interior geometry, Phys. Lett. B {\bf 797}, 134908 (2019).

\bibitem{Bojobh} M. Bojowald, Black-hole models in loop quantum gravity, Universe {\bf 6}, 125 (2020).

\bibitem{GOPbheff} R. Gambini, J. Olmedo, and J. Pullin, Spherically symmetric loop quantum gravity: Analysis of improved dynamics, Class. Quantum Grav. {\bf 37}, 205012 (2020).

\bibitem{Edbh1} J.G. Kelly, R. Santacruz, and E. Wilson-Ewing, Effective loop quantum gravity framework for vacuum spherically symmetric spacetimes, Phys. Rev. D {\bf 102}, 106024 (2020).

\bibitem{Edbh2} J.G. Kelly, R. Santacruz, and E. Wilson-Ewing, Black hole collapse and bounce in effective loop quantum gravity, Class. Quantum Grav. {\bf 38}, 04LT01 (2021).

\bibitem{AOS1} A. Ashtekar, J. Olmedo, and P. Singh, Quantum transfiguration of Kruskal black holes, Phys. Rev. Lett. {\bf 121}, 241301 (2018).

\bibitem{AOS2} A. Ashtekar, J. Olmedo, and P. Singh, Quantum extension of the Kruskal spacetime, Phys. Rev. D {\bf 98}, 126003 (2018).

\bibitem{AOg} A. Ashtekar and J. Olmedo, Properties of a recent quantum extension of the Kruskal geometry, Int. J. Mod. Phys. D {\bf 29}, 2050076 (2020).

\bibitem{gang} A. Ashtekar, J. Lewandowski, D. Marolf, J. Mour\~{a}o, and T. Thiemann, Quantization of diffeomorphism invariant theories of connections with local degrees of freedom, J. Math. Phys. {\bf 36}, 6456 (1995).

\bibitem{Thiemannreg} T. Thiemann, Anomaly-free formulation of non-perturbative, four-dimensional Lorentzian quantum gravity, Phys. Lett. B {\bf 380}, 257 (1996).

\bibitem{ABG} A. García-Quismondo, B. Elizaga Navascu\'es, and G.A. Mena Marugán, The space of solutions of the Ashtekar-Olmedo-Singh effective black hole model, arXiv:2207.04677.

\bibitem{norbert1} N. Bodendorfer, F.M. Mele, and J. M\"{u}nch, A note on the Hamiltonian as a polymerization parameter, Class. Quantum Grav. {\bf 36}, 187001 (2019).

\bibitem{norbert2} N. Bodendorfer, F.M. Mele, and J. M\"{u}nch, Effective quantum extended spacetime of polymer Schwarzschild black hole, Class. Quantum Grav. {\bf 36}, 195015 (2019).

\bibitem{norbert3} N. Bodendorfer, F.M. Mele, and J. M\"{u}nch, Mass and horizon Dirac observables in effective models of quantum black-to-white hole transition, Class. Quantum Grav. {\bf 38}, 095002 (2021).

\bibitem{norbert4} N. Bodendorfer, F.M. Mele, and J. M\"{u}nch, $(b,v)$-type variables for black to white hole transitions in effective loop quantum gravity, Phys. Lett. B {\bf 819}, 136390 (2021).

\bibitem{GQMbh} A. Garc\'ia-Quismondo and G.A. Mena Marug\'an, Exploring alternatives to the Hamiltonian calculation of the Ashtekar-Olmedo-Singh black hole solution, Front. Astron. Space Sci. {\bf 8}, 701723 (2021).

\bibitem{AG2} A. García-Quismondo and G.A. Mena Marugán, Two-time alternative to the Ashtekar-Olmedo-Singh black hole interior, Phys. Rev. D {\bf 106}, 023532 (2022).

\bibitem{Mariam} M. Bouhmadi-L\'opez, S. Brahma, C.-Y. Chen, P. Chen, and D.-h- Yeom, Asymptotic non-flatness of an effective black hole model based on loop quantum gravity, Phys. Dark Univ. {\bf 30}, 100701 (2020).

\bibitem{Bojocov} M. Bojowald, No-go result for covariance in models of loop quantum gravity, Phys. Rev. D {\bf 102}, 046006 (2020).

\bibitem{APS} A. Ashtekar, T. Pawlowski, and P. Singh, Quantum nature of the big bang: Improved dynamics, Phys. Rev. D {\bf 74}, 084003 (2006).

\bibitem{KS1} R. Kantowski and R.K. Sachs, Some spatially inhomogeneous dust models, J. Math. Phys. {\bf 7}, 443 (1966).

\bibitem{KS2} E. Weber, Kantowski–Sachs cosmological models as big‐bang models, J. Math. Phys. {\bf 26}, 1308 (1985).

\bibitem{KS3} K.S. Adhav, V.G. Mete, A.S. Nimkar, and A.M. Pund, Kantowski-Sachs cosmological model in general theory of relativity, Int. J. Theor. Phys. {\bf 47}, 2314 (2008).

\bibitem{Taveras} V. Taveras, LQC corrections to the Friedmann equations for a universe with a free scalar field, Phys. Rev. D {\bf 78}, 064072 (2008).

\bibitem{Dirac} P.A.M. Dirac, Lectures on Quantum Mechanics (Belfer Graduate School of Science, Yeshiva University, New York, 1964).

\bibitem{ze} J.M. Velhinho, The quantum configuration space of loop quantum cosmology, Class. Quantum Grav. {\bf 24}, 3745 (2007).

\bibitem{mmo} M. Mart\'{\i}n-Benito, G.A. Mena Marug\'an, and J. Olmedo, Further improvements in the understanding of isotropic loop quantum cosmology, Phys. Rev. D {\bf 80}, 104015 (2009).

\bibitem{simon} M. Reed and B. Simon, Functional Analysis (Elsevier, San Diego, 1980).

\bibitem{galindo} A. Galindo and P. Pascual, Quantum Mechanics I (Springer-Verlag, Berlin, 1990).

\bibitem{mop} G.A. Mena Marug\'an, J. Olmedo, and T. Paw\l owski, Prescriptions in loop quantum cosmology: A comparative analysis, Phys. Rev. D {\bf 84}, 064012 (2011).

\bibitem{measure} P. Billingsley, Probability and Measure (Third ed., John Wiley \& Sons, New York, 1995).

\bibitem{Mariamquasi} M. Bouhmadi-L\'opez, S. Brahma, C.Y. Chen, P. Chen, and D. Yeom, A consistent model of non-singular Schwarzschild black hole in loop quantum gravity and its quasinormal modes, JCAP {\bf 07}, 066 (2020).

\bibitem{Greenquasi} R.G. Daghigh, M.D. Green, and G. Kunstatter, Scalar perturbations and stability of a loop quantum corrected Kruskal black hole, Phys. Rev. D {\bf 103}, 084031
(2021).

\bibitem{Javiquasi} D. del-Corral and J. Olmedo, Breaking of isospectrality of quasinormal modes in nonrotating loop quantum gravity black holes, Phys. Rev. D {\bf 105}, 064053 (2022).

\bibitem{massmax} B. Elizaga Navascu\'es, D. Mart\'in de Blas, and G.A. Mena Marug\'an, Time-dependent mass of cosmological perturbations in the hybrid and dressed metric approaches to loop quantum cosmology, Phys. Rev. D {\bf 97}, 043523 (2018).

\end{thebibliography}
\end{document}